\documentclass[prd,notitlepage,superscriptaddress,nofootinbib,preprintnumbers]{revtex4-2}

\usepackage{amsmath}
\usepackage{amsfonts}
\usepackage{extarrows}
\usepackage{graphicx}
\usepackage[utf8]{inputenc}
\usepackage{slashed}
\usepackage{xcolor}
\usepackage{multirow}
\usepackage{braket}
\usepackage{bbold}
\usepackage{placeins}
\usepackage{subcaption}
\usepackage{ragged2e}
\usepackage{mathtools}
\usepackage{upgreek}
\DeclareCaptionJustification{justified}{\justifying}
\usepackage{hyperref}
\hypersetup{colorlinks, linkcolor = [rgb]{0,0.0,0.75}, citecolor = [rgb]{0,0.0,0.75}, urlcolor = [rgb]{0,0.0,0.75}}

\makeatletter
\g@addto@macro\bfseries{\boldmath}
\makeatother

\DeclareMathOperator{\Tr}{Tr}

\DeclareMathOperator{\fm}{{\rm fm}}
\DeclareMathOperator{\mev}{{\rm MeV}}
\DeclareMathOperator{\gev}{{\rm GeV}}
\newcommand{\MSbar}{\overline{\text{MS}}}
\newcommand{\MMS}{\text{M}\MSbar}

\newcommand{\quoteh}[1]{``{#1}''}

\begin{document}


\title{Flavor decomposition of the nucleon unpolarized, helicity and transversity parton distribution functions from lattice QCD simulations}

\author{Constantia~Alexandrou}
\affiliation{Department of Physics, University of Cyprus, P.O.\ Box 20537, 1678 Nicosia, Cyprus}
\affiliation{Computation-based Science and Technology Research Center, The Cyprus Institute, 20 Kavafi Str., Nicosia 2121, Cyprus}
\author{Martha~Constantinou}
\affiliation{Temple University, 1925 N.\ 12th Street, Philadelphia, PA 19122-1801, USA}
\author{Kyriakos~Hadjiyiannakou}
\affiliation{Department of Physics, University of Cyprus, P.O.\ Box 20537, 1678 Nicosia, Cyprus}
\affiliation{Computation-based Science and Technology Research Center, The Cyprus Institute, 20 Kavafi Str., Nicosia 2121, Cyprus}
\author{Karl~Jansen}
\affiliation{NIC, Deutsches Elektronen-Synchrotron, 15738 Zeuthen, Germany}
\author{Floriano~Manigrasso}
\affiliation{Department of Physics, University of Cyprus, P.O.\ Box 20537, 1678 Nicosia, Cyprus}
\affiliation{Institut für Physik, Humboldt-Universität zu Berlin, Newtonstr.\ 15, 12489 Berlin, Germany}
\affiliation{Dipartimento di Fisica, Università di Roma ``Tor Vergata'', Via della Ricerca Scientifica 1, 00133 Rome, Italy\\
\vspace*{-0.25cm}
\centerline{\includegraphics[scale=0.18]{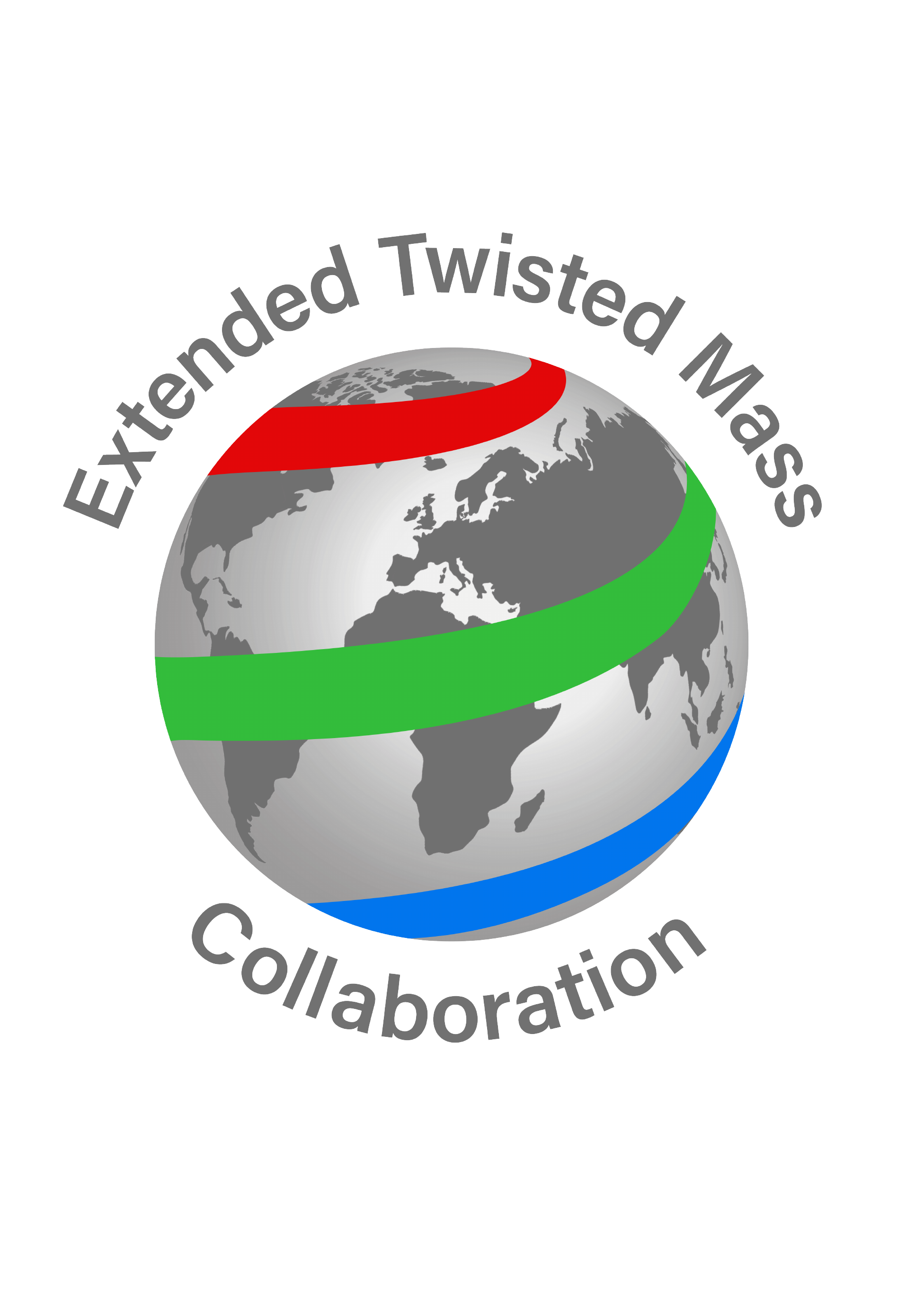}}}
\date{\today}

\begin{abstract}
\vspace*{-0.2cm} 

We present results on the quark unpolarized, helicity and transversity parton distributions functions of the nucleon. We use the quasi-parton distribution approach within the lattice QCD framework and perform the computation using an ensemble of twisted mass fermions  with the strange and charm quark masses tuned to approximately their physical values and  light quark masses giving pion mass of 260~MeV. We use hierarchical probing to evaluate the disconnected quark loops. We discuss identification of ground state dominance,  the Fourier transform procedure and convergence with the momentum boost. We find non-zero results for the disconnected isoscalar and strange quark distributions. The determination of the quark parton distribution and in particular the strange quark contributions that are poorly known provide valuable input to the structure of the nucleon.   
\end{abstract}

\maketitle

\section{Introduction}

Parton distribution functions (PDFs) are the foundation for understanding the structure of hadrons in terms of their partonic content. Together with the generalized parton distributions (GPDs) and the transverse-momentum dependent distributions (TMDs) form a set of quantities that are needed for the mapping of hadrons, in the coordinate and momentum space. The most well-constraint distribution functions are the PDFs, which depend only on the momentum fraction carried by the partons. These can be obtained from a number of scattering processes and have a wide kinematical coverage (see, e.g., Ref.~\cite{Ethier:2020way,Nocera:2014gqa}). Furthermore, the individual-flavor contributions are also studied in phenomenological fits for the colinear PDFs. 

The present calculation is motivated by the fact that not all PDFs are well-constrained from global analyses. The number of available experimental data sets in the case of the transversity is less by ${\cal O}(10)$ compared to the helicity, and ${\cal O}(100)$ compared to the unpolarized PDFs. In addition, isolating the strange-quark PDF from the down-quark PDFs can be challenging, as most of the high-energy processes cannot differentiate between the two flavors. For example, there is a disagreement on the sign of the strange-quark helicity PDF, $\Delta s (x) + \Delta \bar{s} (x)$, from analysis of polarized inclusive deep inelastic scattering~\cite{Leader:2006xc,Leader:2014uua} and global analyses of inclusive and semi-inclusive deep inelastic scattering data sets~\cite{deFlorian:2009vb,Leader:2010rb,Arbabifar:2013tma,Leader:2011tm}. The large uncertainties in the strange PDFs have an effect on other quantities, such as the $W$-boson mass and the determination of the CKM matrix element $V_{cs}$~\cite{Aaboud:2017svj,Alekhin:2017olj}. Therefore, calculations of the individual-quark PDFs from lattice QCD can, eventually, be used as input in analysis requiring knowledge of PDFs. 

The flavor decomposition of proton charges and form factors had been under investigation in the last few years with calculations of disconnected diagrams using ensembles at or near the physical values for the quark masses(\textit{physical point})~\cite{Abdel-Rehim:2016won,Alexandrou:2017qyt,Alexandrou:2017hac,Alexandrou:2017ypw,Alexandrou:2017oeh,Alexandrou:2018zdf,Alexandrou:2018sjm,Alexandrou:2019brg,Alexandrou:2019olr,Alexandrou:2020sml,Bhattacharya:2015wna,Liang:2019xdx,Djukanovic:2019jtp,Green:2015wqa}.  Such a success in lattice calculations of hadron structure is partly due to available computational resources, but also due to novel methodologies to extract the individual quark Mellin moments and form factors. A notable example is the hierarchical probing~\cite{Stathopoulos:2013aci}, which improves the signal significantly. In this work, we extend  hierarchical probing to non-local operators. Such an approach was shown to be successful in our first calculation on the helicity PDF~\cite{Alexandrou:2020uyt}.  

Matrix elements of non-local operators are of great interest in recent years. These can be related to light-cone PDFs through a factorization and matching procedure. Methods to access the $x$ dependence of PDFs, such as the quasi-PDFs~\cite{Ji:2013dva,Ji:2014gla}, pseudo-ITDs~\cite{Radyushkin:2016hsy}, and current-current correlators~\cite{Ma:2014jla,Ma:2014jga} are now well established. Progress in terms of the renormalizability, renormalization prescription, and factorization of light-cone PDFs has been made, alleviating major sources of systematic uncertainties. For their application in lattice QCD see the recent reviews of Ref.~\cite{Cichy:2018mum,Ji:2020ect,Constantinou:2020pek}. Results from lattice QCD simulations on the $x$-dependence of PDFs are very promising, and therefore, their flavor decomposition is the extension of these investigations. The matching for the singlet case, as well as the mixing with the gluon PDFs have been recently addressed~\cite{Zhang:2018diq,Wang:2019tgg}. In Ref.~\cite{Alexandrou:2020uyt} we presented the first calculation for the helicity PDFs including disconnected contributions and the flavor decomposition for the up-, down- and strange-quark PDFs. Here we extend the calculation to the three types of collinear PDFs, that is the unpolarized, helicity and transversity PDFs. While such calculations are becoming feasible, there are a number of computational challenges before taming the statistical uncertainties. To date, calculations of disconnected contributions for matrix elements of non-local operators at the physical point do not exist. 

The paper is organized as follows: Section~\ref{sec:meth} presents the methodology of extracting the nucleon matrix elements, the renormalization and matching. Section~\ref{sec:latt} focuses on the details of the calculation and the techniques employed. The results for the disconnected and connected matrix elements are presented in Sections~\ref{sec:disc} and \ref{sec:conn_mat_el}, respectively. In Section~\ref{sec:charges}, we extract the vector, axial and tensor charges using the data with zero length for the Wilson line. The PDF reconstruction and flavor decomposition is given in Section~\ref{sec:PDFs}. Finally, we give out conclusions in Section~\ref{sec:concl}.

\FloatBarrier
\section{Methodology}
\label{sec:meth}
\FloatBarrier
\subsection{Nucleon bare matrix elements}\label{sec:theory_matrix_el}

The main component of this study is the calculation of the nucleon matrix elements of non-local operators, that is
\begin{equation}\label{eq:matrix_el}
    h_{\Gamma}^{\Upupsilon}(z,P_3) = \braket{N(P_3)|\overline{\psi}(z)\Upupsilon\, \Gamma \, W(z) \psi(0)|N(P_3)},
\end{equation}
where $| N(P_3) \rangle$ is the nucleon state with momentum boost along the $z$-direction, i.e. $\vec{P}=(0,0,P_3)$.
The fermionic field $\psi(x) \equiv \psi(\vec{x},t)$ can be either the light quark doublet $\psi\equiv (u,d)^T$ or the strange quark field $\psi(x)\equiv s(x)$.
The Wilson line $W(z)$ is constructed in the direction parallel to the nucleon boost $\vec{P}$ and extends from zero length to up to half of the lattice, $L/2$, in both positive and negative directions. The Dirac structure of the operator, $\Gamma$, acts in spin space and depends on the type of the collinear PDF under study. Without loss of generality, one can take that the momentum boost is in the $z$-direction ($k$). Based on this, we employ the following $\Gamma$ matrices: 
\begin{itemize}
    \item $\Gamma=\gamma^0$ for the unpolarized distribution $q(x)$;
    \item $\Gamma = \gamma^5 \gamma^3$ for the helicity distribution $\Delta q(x)$;
    \item $\Gamma = \sigma_{3j}$ with $j\neq 3$ for the transversity distribution $\delta q(x)$.
\end{itemize}
We calculate both the isovector and isoscalar quark contributions to Eq.~(\ref{eq:matrix_el}), and, therefore, we introduce the superscript $\Upupsilon$ for the matrix elements. The matrix $\Upupsilon$ acts on the light quark sector and takes the value $\tau^3\,(\mathbb{1})$ for the isovector (isoscalar) distribution, where $\tau^3={\rm diag}(1,-1)$, is the third Pauli matrix.

The matrix elements are computed from the ratio of three- and two-point functions defined as
\begin{equation} \label{eq:two-three-point}
\begin{split}
C_{2pt}(\vec{P};t_s,0) &= \mathcal{P}_{\alpha\beta} \sum\limits_{\vec{x}}e^{-i\vec{P}\cdot\vec{x}} \braket{\Omega|N_\alpha(\vec{x},t_s)\overline{N}_\beta(\vec{0},0)|\Omega} \\
C_{3pt}(\vec{P};t_{\rm ins} ;t_s,0) &= \tilde{\mathcal{P}}_{\alpha\beta} \sum\limits_{\vec{x},\vec{y}}e^{-i\vec{P}\cdot \vec{x}}\braket{\Omega|N_\alpha(\vec{x},t_s)\mathcal{O}(\vec{y},t_{\rm ins};z)\overline{N}_\beta(\vec{0},0)|\Omega},
\end{split}
\end{equation}
where $t_s$ is the source-sink separation, and $t_{\rm ins}$ the insertion time of the three-point function. We use the proton interpolating field $N_\alpha=\varepsilon^{abc}u_\alpha^a(x)\left( d^{bT}(x)\mathcal{C}\gamma^5u^c(x)\right)$ with $\mathcal{C}=\gamma^0\gamma^2$.
The three-point function projector depends on the operator under study and can be found in Table~\ref{table:parity_proj} and for the two-point function we use ${\cal P}=(1 \pm \gamma_0)/2$. To increase the number of measurements for the disconnected diagrams, we average the three- and two- point functions over plus and minus parity projectors.

\begin{table}[h]
\begin{tabular}{c c c} 
 \hline
 \hline
 \noalign{\vskip 0.1cm}    
 PDF & $\Gamma$ & $\tilde{\mathcal{P}}_{\alpha\beta}$ \\ [0.5ex]
 \hline\hline
  \noalign{\vskip 0.2cm}   
 Unpolarized & $\gamma^0$ & $\frac{\mathbf{1}\pm\gamma^0}{2}$  \\ 
  \noalign{\vskip 0.2cm}  
 Helicity & $\gamma^5\gamma^3$ & $i\gamma^3\gamma^5\left(\frac{\mathbf{1}\pm\gamma^0}{2}\right)$  \\ 
  \noalign{\vskip 0.2cm}  
 Transversity & $\sigma_{3j}$ & $i\gamma^5\gamma^i\left(\frac{\mathbf{1}\pm\gamma^0}{2}\right),\;i\neq j$  \\ 
  \noalign{\vskip 0.1cm}  
 \hline
 \hline
\end{tabular}
\caption{ List of parity projectors and insertions for each colinear PDFs. The nucleon momentum boost is assumed to be in the $z$-direction, $P=(0,0,P_3)$.}
\label{table:parity_proj}
\end{table}

\begin{figure}
    \centering
    \begin{subfigure}[t]{0.30\textwidth}
        \centering
        \includegraphics[width=1.\linewidth]{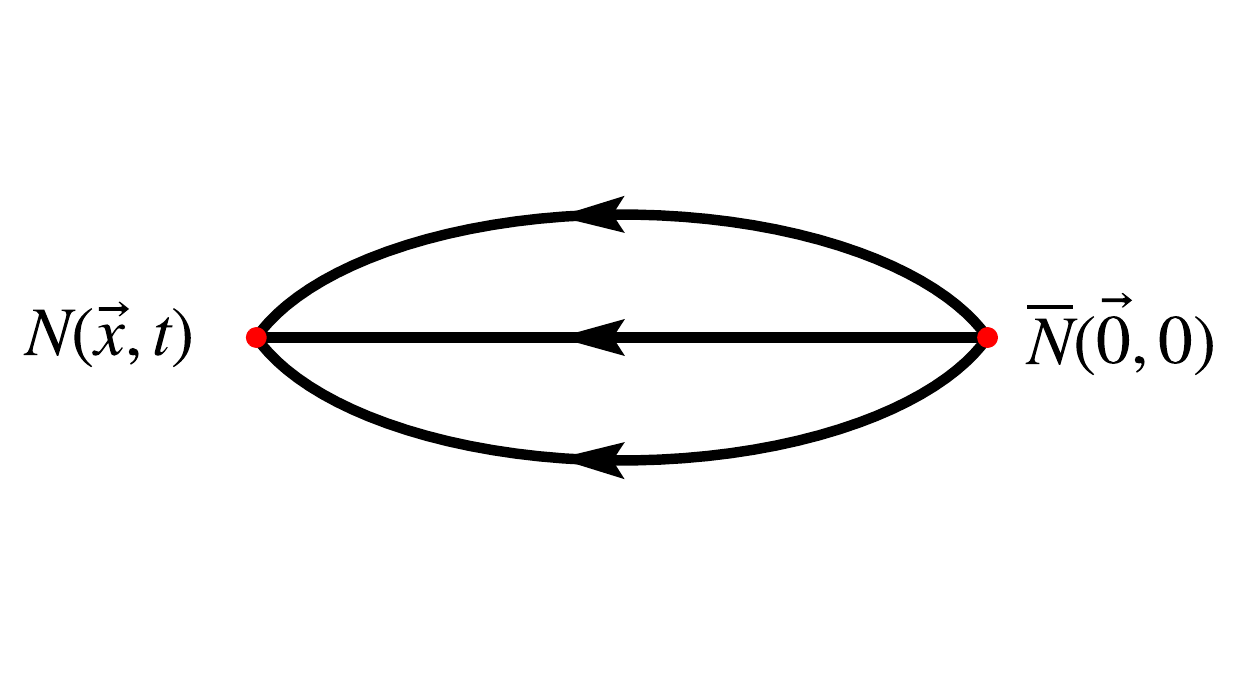} 
        \caption{Nucleon two-point function.} \label{fig:timing1}
    \end{subfigure}
    \hfill
    \begin{subfigure}[t]{0.32\textwidth}
        \centering
        \includegraphics[width=1.05\linewidth]{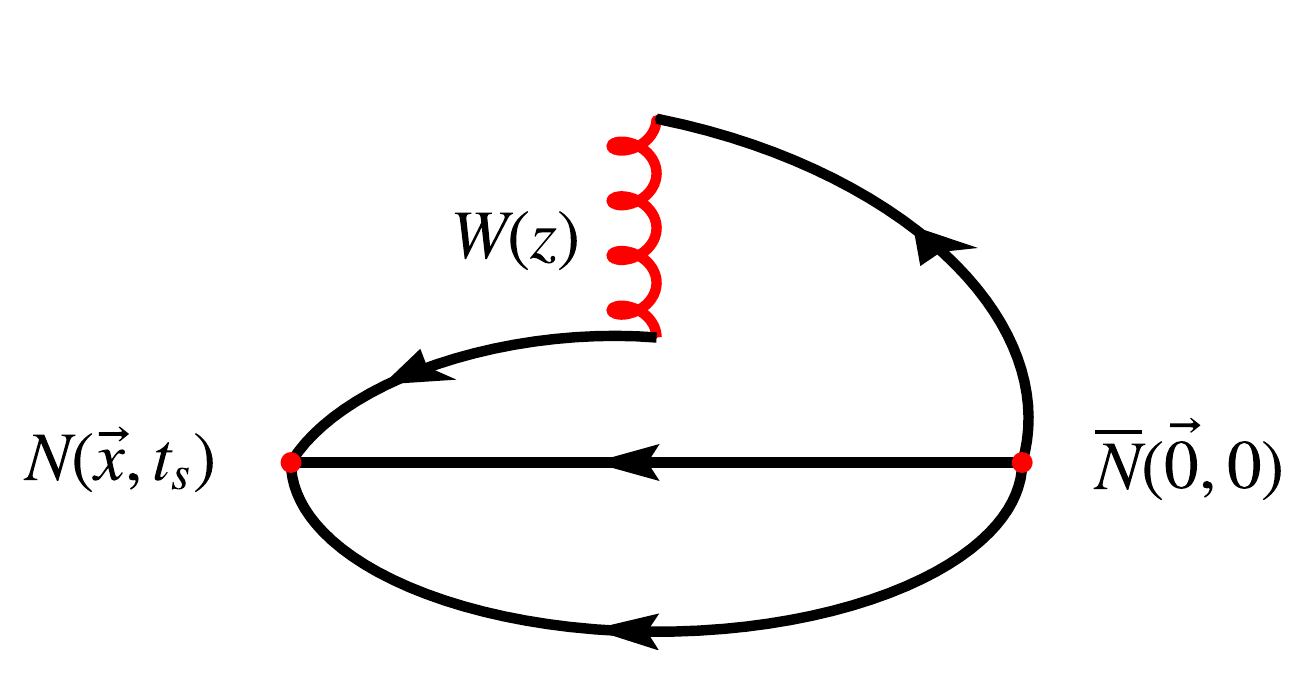} 
        \caption{Nucleon connected three-point function.} \label{fig:diag_threep_conn}
    \end{subfigure}
    \hfill
    \begin{subfigure}[t]{0.3\textwidth}
    \centering
        \includegraphics[width=1\linewidth]{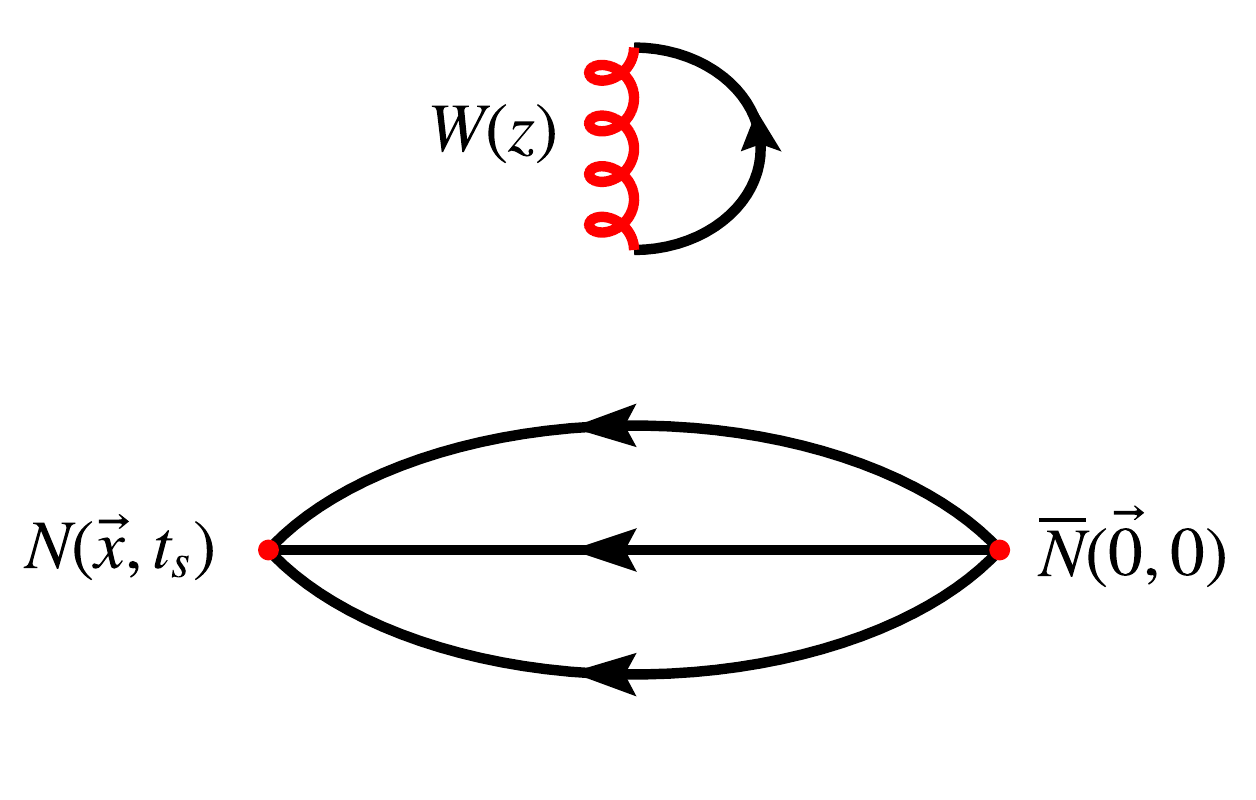} 
        \caption{Nucleon disconnected three-point function.} \label{fig:diag_threep_disc}
    \end{subfigure}
    \caption{Schematic representation of the two- and three-point functions. The time $t_s\,(t)$ indicates the source-sink separation for the three(two)-point function. The solid lines correspond to quark propagators, while the curly lines represent the Wilson lines of length $z$. }
    \label{fig:diagrams}
\end{figure}

The operator $\mathcal{O}$ is defined as
\begin{equation} \label{eq:operator_insertion}
    \mathcal{O}(\vec{y},t_{\rm ins};z) = \bar{\psi}(\vec{y}+z\hat{z},t_{\rm ins})\Upupsilon\,\Gamma\, W(\vec{y}+z\hat{z},\vec{y})\psi(\vec{y},\tau)\,,
\end{equation}
and is inserted in the three-point function of Eq.~\eqref{eq:two-three-point}. The Wick contractions lead to two topologically different diagrams, as shown in Fig.~\ref{fig:diag_threep_conn} and \ref{fig:diag_threep_disc}. In Fig.~\ref{fig:timing1}, we also show pictorially the two-point function of Eq.~\eqref{eq:two-three-point}. For the case that the fermionic field in Eq.~\eqref{eq:operator_insertion} is $\psi=(u,d)^T$ and $\Upupsilon=\tau^3$, we obtain the matrix elements for the isovector distribution $u-d$, which receive contribution from the connected diagram only (Fig.~\ref{fig:diag_threep_conn}). However, in the case where $\Upupsilon=\mathbb{1}$, the three-point function takes contributions from both connected and disconnected diagrams. For the nucleon, the strange-quark contribution comes exclusively from the disconnected diagram. We emphasize that, disconnected contributions have a considerably smaller signal-to-noise ratio compared to the connected ones and their evaluation requires the use of stochastic and gauge noise reduction techniques described in detail in Sec.~\ref{sec:disc_contr}. 

The ratio of three- over two-point functions becomes,
\begin{equation}\label{eq:ratio_3pt_2pt}
 \frac{\braket{C_{3pt}(P_3;t_{\rm ins};t_s,0}}{\braket{C_{2pt}(P_3;t_s,0)}}   \stackrel{0\ll t_{\rm ins} \ll t_s}{=}  \left(h_{\Gamma}^{\Upupsilon}\right)^{\rm bare}(z,P_3)\,\,,
\end{equation}
and is used to obtain the matrix elements of Eq.~\eqref{eq:matrix_el}. This relation is meaningful for the ground-state contribution. To isolate the latter, we apply constant (plateau) fits in a range where the operator insertion time is large enough, and away from the source and sink. Besides the plateau fit method, we employ different techniques allowing the extraction of the nucleon matrix element, as described in Sec.~\ref{sec:excited_states}.

\FloatBarrier
\subsection{Non-perturbative renormalization}

The bare matrix elements of Eq.~(\ref{eq:matrix_el}) must be renormalized in coordinate space prior to obtaining the quasi-PDFs, which are defined in the momentum space (see Sec.~\ref{sec:quasiPDF}). The renormalization of both the non-singlet and singlet quantities is multiplicative. In this work, we use the non-singlet renormalization function, as the difference with the singlet is expected to be small~\cite{Constantinou:2016ieh}, which was demonstrated numerically for local operators~\cite{Alexandrou:2020sml,Alexandrou:2019brg}. The mixing between the unpolarized and helicity singlet-quark PDFs with the gluon PDFs arises at the matching level because there is no additional non-local ultraviolet divergence in the quasi-PDF~\cite{Green:2017xeu,Zhang:2018diq,Wang:2019tgg}. 

To renormalize the matrix elements we apply the regularization independent (RI$'$) scheme, and  use the momentum source method~\cite{Gockeler:1998ye} that offers high statistical accuracy. More details on the setup can be found in Refs.~\cite{Alexandrou:2015sea,Alexandrou:2019lfo}. In Refs.~\cite{Constantinou:2017sej,Alexandrou:2017huk}, we proposed an extension of the renormalization prescription to include non-local operators, which we also follow in this work. The conditions for the renormalization functions of the non-local operator, $Z_{\Gamma}$, and the quark field, $Z_q$, are
\begin{eqnarray}
\label{renorm}
\left(Z_q^{\rm impr}(\mu_0) \right)^{-1}\, Z_{\Gamma}(z,\mu_0)  \, {\rm Tr} \left[{\cal V}_{\Gamma}(p,z) \, \slashed{p} \right] \Bigr|_{p^2{=} \mu_0^2} =
{\rm Tr} \left[{\cal V}_{\Gamma}^{{\rm Born}}(p,z)\, \slashed{p} \right] \Bigr|_{p^2{=} \mu_0^2}
\, , \\[3ex]
  Z^{\rm impr}_q (\mu_0) = \left(\frac{1}{12} {\rm Tr} \left[(S(p))^{-1}\, S^{\rm Born}(p)\right]  - dZ^\infty_q(p) \right) \Bigr|_{p^2= \mu_0^2}\,. \qquad
\end{eqnarray}
${\cal V}(p,z)$ ($S(p)$) is the amputated vertex function of the operator (fermion propagator) and $S^{{\rm Born}}(p)$ is the tree-level of the propagator. These conditions are applied at each value of $z$ separately. We improve $Z_q$, and consequently $Z_{\Gamma}$, by subtracting lattice artifacts calculated to one-loop level in perturbation theory and to all orders in the lattice spacing, $dZ^\infty_q(p)$. The details of the calculation can be found in Ref.~\cite{Alexandrou:2015sea}.

The RI-type schemes are mass-independent, and therefore $Z_{\Gamma}$ is calculated at several ensembles with different values for the quark masses. Eventually, a chiral extrapolation is applied to remove residual artifacts related to the quark mass. Here we use five ensembles that are generated at different pion masses in the range of 350 MeV - 520~MeV with a lattice volume of $24^3 \times 48$. For the chiral extrapolation to be meaningful, all quark masses should be degenerate. Therefore, we use $N_f=4$ ensembles generated by the Extended Twisted Mass collaboration (ETMC) that are dedicated to the renormalization program. These ensembles have the same lattice spacing and action parameters as the $N_f=2+1+1$ ensemble used for the production of the nucleon matrix elements. 

$Z_{\Gamma}$ depends on the RI renormalization scale $\mu_0$, and it will be converted to $\MSbar$ and evolved at a scale of choice. To reliably perform this procedure we use several values of $\mu_0$, and the conversion and evolution formulas is applied on $Z_{\Gamma}$ obtained at each scale. We choose the initial scale $\mu_0$ such that discretization effects are small~\cite{Alexandrou:2015sea}. In particular, the 4-vector momentum $p$, which is set equal to $\mu_0$, has the same spatial components: $p=(p_0,p_1,p_1,p_1)$. The values of $p_0$ and $p_1$ are chosen such that the ratio $\frac{p^4}{(p^2)^2}$ is less than 0.28, as suggested in Ref.~\cite{Constantinou:2010gr}. The values of $a\,\mu_0$  cover the range $[1,5]$.
For each $\mu_0$ value, we apply a chiral extrapolation using the fit
\begin{equation}
\label{eq:Zchiral_fit}
Z^{\rm RI}_{\Gamma}(z, \mu_0,m_\pi) = {Z}^{\rm RI}_{{\Gamma},0}(z, \mu_0) + m_\pi^2 \,{Z}^{\rm RI}_{{\Gamma},1}(z, \mu_0) \,,
\end{equation}
to extract the mass-independent ${Z}^{\rm RI}_{{\Gamma},0}(z, \mu_0)$ at each value of the initial scale. ${Z}^{\rm RI}_{{\Gamma},0}(z, \mu_0)$ is converted to the $\overline{\rm MS}$ scheme and evolved to $\mu{=}2$ GeV ($\mu{=}\sqrt{2}$ GeV) using the results of Ref.~\cite{Constantinou:2017sej} for the unpolarized and helicity (transversity) PDFs. The conversion and evolution depends on both the initial scale $\mu_0$ and the final scale in the $\MSbar$. The appropriate expressions have been obtained to one-loop perturbation theory in dimensional regularization. Therefore, there is residual dependence on the initial scale $\mu_0$, which is eliminated by taking the limit $(a\,\mu_0)^2 \to 0$ using a linear fit on the data in the region $(a\,\mu_0)^2 \in [1-2.6]$. 

The last step of the renormalization program is the conversion to a modified $\MSbar$ scheme ($\MMS$), developed in Ref.~\cite{Alexandrou:2019lfo}, and is given by
\begin{equation}
\label{eq:ZMMS}
\displaystyle {\cal Z}^{\MMS}_{{\Gamma},0}(z,\bar\mu) = {\cal Z}^{\MSbar}_{{\Gamma},0}(z,\bar\mu)\,  {\cal C}^{\MSbar,{\rm M\overline{MS}}}\,,
\end{equation}
where
\begin{eqnarray}
\label{eq:CMStoMMS}
\hspace*{-0.45cm}
{\cal C}_\Gamma^{\overline{\rm MS}, {\rm M\overline{MS}}} {=} 1+\frac{C_F g^2}{16\pi^2} \Bigg[
&{\,}&\hspace*{-0.15cm} e^{(1)}_\Gamma +  e^{(2)}_\Gamma \ln \left(\frac{\bar\mu ^2}{4 \mu_F^2}\right)  
+  e^{(3)}_\Gamma   \left(\frac{i \pi  \left| \mu_F z\right| }{2 \mu_F z}-\ln (\left|
   \mu_F z\right| )-\text{Ci}(\mu_F z)-i \text{Si}(\mu_F z)+\ln
   (\mu_F z)\right)\nonumber \\
 &+&   e^{(4)}_\Gamma  \left(e^{i \mu_F z} (2 \text{Ei}(-i \mu_F
   z)+i \pi  \text{sgn}(\mu_F z)-\ln (-i \mu_F z)+\ln (i \mu_F z))\right)\Bigg]\,.
\end{eqnarray}
This scheme was introduced to satisfy particle number conservation. In Ref.~\cite{Alexandrou:2019lfo} we showed that the difference between $\MSbar$ and $\MMS$ is numerically very small, but brings the PDFs closer to the phenomenological ones. In Eq.~(\ref{eq:CMStoMMS}) $\mu_F$ is the factorization scale set equal to the $\MSbar$ scale. ${\rm Ci}$, ${\rm Si}$, ${\rm Ei}$ and ${\rm sgn}$ are the special functions cosine integral, sine integral, exponential integral, and sign function, respectively. The coefficients $e^{(i)}_\Gamma$ depend on the operator: $\{e_\Gamma^{(1)},\,e_\Gamma^{(2)},\,e_\Gamma^{(3)},\,e_\Gamma^{(4)} \}$ is $\{-5,\,-3,\,+3,\,-3/2\}$, $\{-7,\,-3,\,+3,\,-3/2\}$, $\{-4,\,-4,\,+4,\,-4/2\}$, for the vector, axial and tensor operator, respectively. 

Due to the presence of the Wilson line, both the matrix elements and renormalization functions are complex functions. As a consequence, a complex multiplication is required to extract the renormalized matrix element, that is
\begin{eqnarray}
\label{eq:renorm}
 h_{\Gamma}^{\Upupsilon} = Z^{\MMS}_\Gamma \cdot \left(h_{\Gamma}^{\Upupsilon}\right)^{\rm bare} &=& \, \left({\rm Re}[Z^{\MMS}_\Gamma] \, {\rm Re}[\left(h_{\Gamma}^{\Upupsilon}\right)^{\rm bare}] - {\rm Im}[Z^{\MMS}_\Gamma] \, {\rm Im}[\left(h_{\Gamma}^{\Upupsilon}\right)^{\rm bare}] \right) \nonumber \\[0.5ex]
 &+& i \left({\rm Re}[Z^{\MMS}_\Gamma] \, {\rm Im}[\left(h_{\Gamma}^{\Upupsilon}\right)^{\rm bare}] + {\rm Im}[Z^{\MMS}_\Gamma] \, {\rm Re}[\left(h_{\Gamma}^{\Upupsilon}\right)^{\rm bare}] \right)\,.
\end{eqnarray}
For simplicity in the notation, the dependence on $z$, $P_3$, scheme and scale is implied. As can be seen, the real (imaginary) part of the renormalized matrix elements are not simple multiples of the real (imaginary) part of the bare matrix element. Therefore, controlling systematic uncertainties in the renormalization is an important aspect of the calculation.

\FloatBarrier
\subsection{Quasi-PDFs and matching to light-cone PDFs}
\label{sec:quasiPDF}

Quasi-PDFs are defined as the Fourier transform of the renormalized nucleon matrix elements in Eq.~\eqref{eq:ratio_3pt_2pt} with respect to the Wilson line length $z$
\begin{equation}\label{eq:quasi-PDF}
    \tilde{q}(x,P_3)=\int\limits_{-\infty}^{\infty} \frac{dz}{4\pi} e^{-i x z P_3} \, h_{\Gamma}^{\Upupsilon}(z,P_3)\,.
\end{equation}
 Note that the renormalized matrix elements, $ h_{\Gamma}^{\Upupsilon}$, depend on the scheme and scale, which also propagates to $\tilde{q}(x,P_3)$. For simplicity in the notation, this dependence is implied. As mentioned in the previous paragraph, the matrix elements are renormalized in the $\MMS$ scheme and evolved to 2~GeV ($\sqrt{2}$ GeV) for the unpolarized and helicity (transversity) PDFs.

On the lattice, we can only evaluate the matrix elements for discrete and finite values of $z$. Therefore, the integral of Eq.~\eqref{eq:quasi-PDF} is replaced by a discrete sum over a finite number of Wilson line lengths 
\begin{equation}\label{eq:quasi-PDF-discrete}
    \tilde{q}(x,P_3)=\sum\limits_{-z_{\rm max}}^{z_{\rm max}} \frac{dz}{4\pi} e^{-ixzP_3} \, h_{\Gamma}^{\Upupsilon}(z,P_3)\,,
\end{equation}
where $dz/a=1$. 
For the summation in Eq.~(\ref{eq:quasi-PDF-discrete}) to accurately reproduce Eq.~(\ref{eq:quasi-PDF}), both the real and imaginary parts of the matrix element should be zero beyond $z_{\rm max}$. Practically, this is not always possible due to the finite momentum boost and limited volume in the lattice formulation. The choice of the cutoff $z_{\rm max}$, which is anyway limited up to $L/2$, requires an extensive study. We note that systematic effects related to the reconstruction of the PDFs is operator dependent, as each matrix element may have different large-$z$ behavior. We will show the results of such analysis in Sec.~\ref{sec:syst_eff_truncation}. 

From the finite-momentum quasi-PDF, it is possible to obtain the light-cone parton distribution function (infinite momentum) through the so-called matching procedure. This is accomplished through a convolution of the quasi distribution with a kernel evaluated in continuum perturbation theory within the large momentum effective theory (LaMET)~\cite{Ji:2020byp,Ji:2020ect}. The matching formula reads
\begin{equation}
\label{eq:matching}
q(x,\mu)=\int_{-\infty}^\infty 
\frac{d\xi}{|\xi|} \, C\left(\xi,\frac{\mu}{x P_3}\right)\, \widetilde{ q}\left(\frac{x}{\xi},\mu,P_3\right)\,,
\end{equation}
and the factorization scale $\mu$ is chosen to be the same as the renormalization scale.
The matching kernel $C$ contains information on $P_3$ which, in principle, is eliminated in $q(x,\mu)$. However, there is residual $P_3$ due to the limitations in accessing large values of momentum from lattice QCD (see, e.g., Ref.~\cite{Alexandrou:2019lfo}) and the matching kernel being available to limited order in perturbation theory. Most of the calculations of the matching kernel have been performed to one-loop level (see, e.g., Refs.~\cite{Xiong:2013bka,Ma:2014jla,Ji:2015qla,Xiong:2015nua,Ma:2017pxb,Wang:2017qyg,Stewart:2017tvs,Izubuchi:2018srq}). Recently, the computation of the kernel $C$ was extended to two loops~\cite{Izubuchi:2018srq,Chen:2020arf,Chen:2020iqi,Chen:2020ody}. In this study, we employ the kernel in the $\MMS$-scheme which is known at one-loop level~\cite{Alexandrou:2019lfo}. This matching kernel relates the quasi-PDFs defined in the $\MMS$ scheme at some scale, to the light-cone PDFs in the $\MSbar$ at the same scale. For the unpolarized and helicity we choose a scale of 2 GeV, while for the transversity we choose $\sqrt{2}$ GeV.

To calculate the anti-quark distributions from $q(x)$, we exploit the crossing relations~\cite{Collins:2011zzd}, that is
\begin{equation}\label{eq:crossing_rels}
    \bar{q}^f(x) = - q^f(-x),\;\;\;\Delta \bar{q}^f(x) =  \Delta q^f(-x),\;\;\;\delta\bar{q}^f(x) = - \delta q^f(-x)\,.
\end{equation}

\FloatBarrier
\section{Lattice setup}
\label{sec:latt}

The computation is performed using one gauge ensemble of $N_f=2+1+1$ clover-improved twisted mass fermions and the Iwasaki improved gluonic action~\cite{Iwasaki:1985we} generated by ETMC~\cite{Alexandrou:2018egz}. The fermionic action of the light quarks in the \quoteh{twisted basis} takes the form
\begin{equation}\label{eq:light_quark_action}
    S_{\rm tm}^l(\chi_l,\overline{\chi}_l,U) = a^4\sum_{x}\overline{\chi}_l(x)\left[D_W[U] + i \mu_l \gamma_5 \tau_3 + m_l + \frac{i}{4}c_{\rm SW} \sigma^{\mu\nu}F^{\mu\nu}[U] \right]\chi_l(x)\,.
\end{equation}
Here, $\chi^T_l(x)=(u,d)$ is the light quark doublet in the twisted basis, $D_W[U]$ is the massless Wilson-Dirac operator and $F_{\mu\nu}[U]$ is the field strength tensor. The last term is weighted by $c_{\rm SW}$, the Sheikoleslami-Wohlert~\cite{Sheikholeslami:1985ij} clover coefficient. The heavy quark twisted mass action is similar to the light quark action in Eq.~\eqref{eq:light_quark_action}. However, it contains an additional term, proportional to the parameter $\mu_\delta$, due to the non-degeneracy of the heavy quarks and reads
\begin{eqnarray}
    S_{\rm tm}^h(\chi_h,\overline{\chi}_h,U) = a^4\sum_{x}\overline{\chi}_h(x)\left[D_W[U] + i \mu_\sigma \gamma_5 \tau_3 + m_h - \mu_{\delta}\tau_1 + \frac{i}{4}c_{\rm SW} \sigma^{\mu\nu}F^{\mu\nu}[U] \right]\chi_h(x)\,,
\end{eqnarray}
where $\chi^T_h(x)=(s,c)$.
Moreover, $\mu_l$ and $\mu_\sigma$ are the twisted mass parameter. The mass terms $m_l$ and $m_h$ are the (untwisted) Wilson quark masses tuned to the critical value $m_{\rm crit}$ (i.e. at \emph{maximal twist}), which ensures automatic $\mathcal{O}(a)$ improvement~\cite{Frezzotti:2003ni} for parity even operators. The equivalent discussion for non-local operators can be found in Refs.~\cite{Green:2017xeu,Chen:2017mie,Green:2020xco,Alexandrou:2020qtt}. 
Fields in the \quoteh{physical basis} can be obtained from the twisted basis through the transformation
\begin{equation}\label{eq:twisted_mass_transf}
    \overline{\psi}(x)\equiv \overline{\chi}(x)e^{i\frac{\alpha}{2}\gamma^5 \tau^3}, \quad \quad \quad \psi(x)\equiv e^{i\frac{\alpha}{2}\gamma^5 \tau^3} \chi(x),
\end{equation}
with $\alpha=\pi/2$ at maximal twist. From now on we will use fields in the physical basis.

The ensemble we use has lattice volume $V=32^3\times 64$, with a lattice spacing of $a=0.0938$ fm. The pion mass is approximately equal to $m_\pi=260\,\mev$ and $m_\pi L \approx 3$. In Table~\ref{Table:params}, one can find the summary of the main parameters characterizing the ensemble. For further details see Ref.~\cite{Alexandrou:2018egz}.

\begin{table}[h!]
\begin{center}
\renewcommand{\arraystretch}{1.5}
\renewcommand{\tabcolsep}{5.5pt}
\begin{tabular}{c|lc}
\hline\hline
\multicolumn{3}{c}{ 
$\beta=1.726$, $c_{\rm SW} = 1.74$, $a=0.0938(3)(2)$~fm}\\
\hline
\multirow{4}{*}{$32^3\times 64$, $L=3.0\fm$\,}  & $\,\,a\mu_l = 0.003$   \\
                                              & $\,\,m_\pi \approx 260 \mev$     \\
                              			      & $\,\,m_\pi L \approx 3$    \\
                        					  & $\,\,m_N = 1.09(6)\,\gev $     \\
\hline\hline
\end{tabular}
\begin{center}
\begin{minipage}{15cm}
\vspace*{-0.45cm}
\hspace*{3cm}
\caption{\small{Parameters of the ensemble used in this work. The nucleon mass $(m_N)$, the pion mass $(m_\pi)$ and the lattice spacing $(a)$ are determined in Ref.~\cite{Alexandrou:2017xwd}.}}
\label{Table:params}
\end{minipage}
\end{center}
\end{center}
\end{table}

\FloatBarrier
\subsection{Numerical methods}

\subsubsection{Connected diagrams}
To improve the overlap between the states generated by the interpolating field $N_\alpha(x)=\varepsilon^{abc}u_\alpha^a(x)\left( d^{bT}(x)\mathcal{C}\gamma^5u^c(x)\right)$ and the proton ground state we employ Gaussian smearing~\cite{Gusken:1989qx,Alexandrou:1992ti}. In addition, we use  APE smearing for the  gauge links that enter the Gaussian smearing. The optimal parameters for the Gaussian and APE smearing techniques, determined in Ref.~\cite{Alexandrou:2021gqw}, are $(\alpha_{\rm G},N_{\rm G})=(4.0,50)$ and $(\alpha_{\rm APE},N_{\rm APE})=(0.5,50)$, respectively. Moreover, to further improve the overlap with the boosted proton ground state, we use momentum smearing~\cite{Bali:2016lva}, as it has been proven to drastically reduce the statistical noise in the matrix elements of boosted hadrons~\cite{Alexandrou:2016jqi}. In Ref.~\cite{Alexandrou:2020zbe} and in this work, the momentum smearing parameter has been tuned to $\xi=0.6$, which minimizes the statistical errors. The momentum smearing operator $\mathcal{S}$ on a quark field $\psi(x)$ reads
\begin{equation}
    \mathcal{S}\psi(x)=\frac{1}{1+6\alpha_G}\left(\psi(x)+\alpha_G \sum\limits_{j=1}^{3}U_j(x)e^{i\xi \vec{P}\cdot \hat{j}}\psi(x+\hat{j})\right),
\end{equation}
where $\xi$ is the momentum smearing parameter and $j$ runs over the spatial directions, with $U_j(x)$ being the link in the spatial $j$-direction. 

To evaluate the connected contributions to the three-point functions, we employ the sequential method~\cite{Martinelli:1988rr} through the sink. Moreover, to further increase the number of measurements, we compute the three-point functions with $N_{\rm src}$ different source positions on each configuration and we boost the nucleon along all the spatial directions and orientations, i.e. $\pm x, \pm y, \pm z$. Indeed, in Ref.~\cite{Alexandrou:2019lfo} it was found that the statistical uncertainty decreases as $1/\sqrt{N_{\rm src} N_{\rm dirs}}$ for all the operators under consideration, with $N_{\rm dirs}=6$ being the number of directions of the nucleon boost. The number of source positions employed depends on the nucleon boost, and is $N_{\rm src}=8$ for the two lowest values of the momentum and $N_{\rm src}=14$ at $P_3=1.24\,\gev$ for the third. The source-sink separation is $t_s=0.94\,\fm$ for  the lowest momentum value, and $t_s=1.13\,\fm$ for the two highest ones. The value employed for $t_s$ at $P_3=1.24\,\gev$ is expected to be large enough to suppress excited-states contamination~\cite{Alexandrou:2019lfo}. In Table~\ref{tab:params_connected} we report the number of measurements for the connected contributions at each value of $P_3$.

\begin{table}[h]
\begin{center}
\renewcommand{\arraystretch}{1.2}
\renewcommand{\tabcolsep}{5.5pt}
\begin{tabular}{c||c c c c}
\hline
\hline
$P_3\,[\gev]$  & $N_{\rm conf}$ & $N_{\rm src}$ & $N_{\rm meas}$ & $t_s\,[{\rm fm}]$  \\
\hline
$0.41$ & $50$ & $8$ & $400$ & $0.94$\\
$0.83$ & $194$ & $8$ & $ 1552$  & $1.13$\\
$1.24$ & $709$ & $14$ & $9926$& $1.13$\\
\hline\hline
\end{tabular}
\caption{Number of measurements, $N_{\rm meas}$, used for the connected diagrams. For each value of $P_3$ we report the number of configurations and source positions employed, as well as the source-sink separation, $t_s$, in physical units.}
\label{tab:params_connected}
\end{center}
\end{table} 

\FloatBarrier
\subsubsection{Disconnected contribution}\label{sec:disc_contr}
The evaluation of the disconnected quark loops with a Wilson line in the boosted frame constitutes the most computationally demanding aspect of this work. The isoscalar three-point function of Eq.~\eqref{eq:two-three-point} ($\tau=\mathbb{1}$ and $\psi(x)=(u(x),d(x))^T$) reads
\begin{equation}\label{eq:3pt_isoscalar_explicit}
    C_{3pt}(\vec{P};t_{\rm ins};t_s,0)=\tilde{\mathcal{P}}_{\alpha \beta} \sum\limits_{\vec{x},\vec{y}}e^{-i\vec{P}\cdot \vec{x}}\braket{\Omega|{\cal J}_N(x)[\overline{u}(y+z)\Gamma W(z)u(y)+\overline{d}(y+z)\Gamma W(z)d(y)]\overline{{\cal J}}_N(0)|\Omega}_{\beta\alpha}.
    \end{equation}
    The three-point function $C_{3pt}$ contains a connected and disconnected part. The latter is given by
\begin{equation}
    C_{3pt}^{\rm disc}(\vec{P};t_{\rm ins};t_s,0)=-\tilde{\mathcal{P}}_{\alpha \beta} \sum\limits_{\vec{x},\vec{y}}e^{-i\vec{P}\cdot {x}}\braket{({\cal J}_N(x)\overline{{\cal J}}_N(0))_{\beta\alpha}\Tr{\left[\left(\mathcal{G}_u(y;y+z)+\mathcal{G}_d(y;y+z)\right)\Gamma W(z)\right]}}\,,
\end{equation}
 where $x=(t_s,\vec{x})$, $y=(t_{\rm ins},\vec{y})$ and $z=(0,0,0,z)$. 
The quantity $\mathcal{G}_{f}(\vec{x},t_x;\vec{y},t_y)$ is the \emph{all-to-all} propagator with quark flavor $f=u,d,s$,  from each lattice point $x$ to any point $y$. Eq.~(\ref{eq:3pt_isoscalar_explicit}) is a correlation of two parts: the nucleon two-point function and the quark loop with a Wilson line. The latter can be written as
\begin{equation}\label{eq:quark_loops}
\begin{split}
    \mathcal{L}^{ \rm u+d}(\tau;z;\Gamma)&=\sum\limits_{\vec{y}}\Tr{\left[\left(\mathcal{G}_u(y;y+z)+\mathcal{G}_d(y;y+z)\right)\Gamma W(z)\right]} \\
    &=\sum\limits_{\vec{y}}\Tr\left[\overline{\psi}(y+z)\Gamma W(z)\psi(y)\right]\,,
\end{split}
\end{equation}
where the trace in the second line is intended over volume, spin and flavor indices.
Due to the presence of the \emph{all-to-all} propagator $\mathcal{G}_{f}(\vec{x},t_x;\vec{y},t_y)$, the exact evaluation of the disconnected contribution in Eq.~\eqref{eq:quark_loops} would require $\approx 10^7$ inversions of the Dirac operator per configuration for the lattice that we are considering. In contrast, stochastic techniques allow to carry out the computation of the disconnected loops with a feasible but yet high computational cost compared to the currently available resources. Stochastic techniques employ noise sources $\xi_r(x)$ that obey two properties
\begin{equation}
    \begin{split}
        \frac{1}{N_r}\sum\limits_{r} \xi_r(x) &= 0 + \mathcal{O}\left(\frac{1}{\sqrt{N_r}}\right)\\
        \frac{1}{N_r}\sum\limits_{r}\xi_r(x)\otimes\xi^*_r(y)&=\delta(x,y)\delta_{\alpha \beta} \delta_{a b}+\mathcal{O}\left(\frac{1}{\sqrt{N_r}}\right)\,,
    \end{split}
\end{equation}
where the product between source vector has to be intended as a tensor product in volume, spin and color subspaces. Given the set of vectors $\xi_r(x)$, the all-to-all propagator can be constructed by solving the equation
\begin{equation}
    M(x,y)\phi_r(y) = \xi_r(x)\,,
\end{equation}
$M(x,y)$ being the Dirac twisted-mass operator. Having the set of solutions ${\phi_r(x)}$, the all-to-all propagator can be estimated via
\begin{equation}
    \mathcal{G}(x;y)=\frac{1}{N_r}\sum\limits_{r}\phi_r(x)\xi_r^\dagger(y) + \mathcal{O}\left(\frac{1}{\sqrt{N_r}}\right)\,.
    \label{Eq:stdLoopDef}
\end{equation}
The number of stochastic vectors required so that the stochastic error becomes comparable to the gauge error should be much smaller than calculating the all-to-all propagator exactly. In addition, exploiting a property of the twisted mass operator, it is possible to design a stochastic algorithm that further reduce the computational cost. Recalling the transformation of Eq.~\eqref{eq:twisted_mass_transf}, the insertion operator $\Gamma$ in twisted basis reads
\begin{equation}
    \Gamma^{\rm tm} \equiv e^{i\frac{\alpha}{2}\gamma^5\tau^3}\Gamma e^{i\frac{\alpha}{2}\gamma^5\tau^3}\,,
\end{equation}
with $\alpha=\pi/2$ at maximal-twist. Depending on the operator $\Gamma$ it is possible to exploit two properties of the twisted-mass operator to evaluate the loop of Eq.~\eqref{eq:quark_loops} with a stochastic technique~\cite{Boucaud:2008xu,Michael:2007vn}:
\begin{enumerate}
    \item If $[\Gamma^{\rm tm},\gamma^5]=0$ for a particular $\Gamma$, then $\tau^3$ appears in the loop of Eq.~\eqref{eq:quark_loops} when expressed in the twisted basis
    \begin{equation}\label{eq:standard_one_end}
    \begin{split}
    \mathcal{L}^{u+d}(\tau;z;\Gamma)&=\mathcal{L}^{u+d}_{\rm tm}(\tau;z;i\gamma^5\tau^3\Gamma)\\
    &=\sum\limits_{\vec{y}}\Tr\left[\left(\mathcal{G}_u^{\rm tm}(y;y+z)-\mathcal{G}^{\rm tm}_d(y;y+z)\right)i\gamma^5 \Gamma W(z,0)\right]\,.
    \end{split}
    \end{equation}
    In this case, we apply the \emph{standard one-end trick}, that exploit the following property of the twisted mass operator
    \begin{equation}
        \mathcal{G}_u^{\rm tm}-\mathcal{G}_{d}^{\rm tm} = -2i\mu (\mathcal{M}_u^\dagger \mathcal{M}_u)^{-1}\gamma^5\,.
    \end{equation}
    The transversity operator $\Gamma=\sigma_{3j}$ belongs  in this category. \\[0.5ex]
    \item If $\{\Gamma^{\rm tm},\gamma^5\}=0$, then the loop in twisted basis possess the same analytical form as in the physical basis
    \begin{equation}\label{eq:general_one_end}
    \begin{split}
        \mathcal{L}^{u+d}(\tau;z;\Gamma)&=\mathcal{L}^{u+d}_{\rm tm}(\tau;z;\Gamma)\\
        &=\sum\limits_{\vec{y}}\Tr\left[\left(\mathcal{G}_u^{\rm tm}(y;y+z)+\mathcal{G}^{\rm tm}_d(y;y+z)\right) \Gamma W(z,0)\right]\,.
        \end{split}        
    \end{equation}
    The quantity of interest can be computed with the  \emph{generalized one-end trick}, exploiting the following property
    \begin{equation}
         \mathcal{G}_u^{\rm tm}+\mathcal{G}_{d}^{\rm tm} = 2\gamma^5 D_W (\mathcal{M}_u^\dagger \mathcal{M}_u)^{-1}\gamma^5\,,
    \end{equation}
    where $D_W$ is the massless Wilson-Dirac clover operator. Note that if the twisted mass parameter becomes very small (close to the physical point) this type of one-end trick is approaching the standard definition in Eq.~\eqref{Eq:stdLoopDef}. The helicity operator $\Gamma=\gamma^5\gamma^3$ and the unpolarized $\Gamma = \gamma^0$ belong to this category.
\end{enumerate}

One of the technical aspects of the calculation is the evaluation of the traces in Eqs.~\eqref{eq:standard_one_end} - \eqref{eq:general_one_end}. With small quark masses, the contribution to the loops coming from the low modes of the spectrum of the Dirac operator may be sizeable, and contributes significantly to the stochastic noise~\cite{Abdel-Rehim:2016pjw}. Therefore, we compute the first $N_{ev}=200$ eigen-pairs $\lambda_j,\ket{v_j}$ of the squared Dirac operator $\mathcal{M}_u\mathcal{M}_u^\dagger$, that allow to reconstruct exactly the low-mode contribution to the disconnected quark loops.
At this stage, stochastic techniques can be employed with the deflated operator to evaluate the high-modes contribution to the traces. To reduce the stochastic noise, we use the \emph{hierarchical probing} algorithm~\cite{Stathopoulos:2013aci}, that allows to reduce the contamination to the trace coming from off-diagonal terms up to a distance $2^k$. This improvement is achieved by partitioning the lattice with $2^{d(k-1)+1}$ Hadamard vectors, where $d=4$ is the number of dimensions of the lattice. Finally, to remove the contamination from off-diagonal terms in spin-color subspaces, we apply full dilution~\cite{Wilcox:1999ab}.
The algorithm employed in the present work has been successfully used in other studies involving the evaluation of disconnected contributions~\cite{Alexandrou:2019brg,Alexandrou:2020sml,Alexandrou:2019olr,Alexandrou:2018sjm}.

For each value of the proton boost $P_3=0.41,0.83,1.24\,\gev$, we evaluated the two-point functions contributing to the disconnected diagram of Eq.~\eqref{eq:3pt_isoscalar_explicit} with $N_{srcs}=200$ source positions (see Sec.~\ref{sec:two_point}). In addition, apart from averaging over all possible directions and orientations of the nucleon boost, we also average over forward and backward projections. In Table~\ref{table:disc_diagrams_stat}, we report the total statistics collected for the disconnected three point correlators. We note that, for the second largest value of the momentum used, namely $P_3=1.24\,\gev$ we use $\approx 10^6$ measurements. In addition, we also compute the matrix elements for $P_3=1.65\,\gev$ and all $\Gamma$ using approximately the same statistics as the previous smaller boost. While this number of statistics is not sufficient to obtain the same statistical accuracy as lower momenta, it allows us to check whether convergence with $P_3$ is reached. 

\begin{table}[h]
\begin{tabular}{c | c c c c c | c c | c} 
 \hline
 \hline
  & \multicolumn{5}{c}{Loops} & \multicolumn{2}{c}{Two-point functions} & \\ [0.5ex]
  \hline\hline
 $P_3\,[\gev]$ & $N_{ \rm ev}$ & $N_{ \rm conf}$ & $N_{ \rm had}$ & $N_{\rm sc}$ & $N_{\rm inv}$ & $N_{\rm srcs}$ & $N_{\rm dir}$ & $N_{\rm meas}$\\
 \hline
    $0.41$ & $200$ & $330$ & $512$ & $12$ & $6144$ & $200$ & $6$ & $396\cdot10^3$\\
    $0.83$ & $200$ & $349$ & $512$ & $12$ & $6144$ & $200$ & $6$ & $418.8\cdot10^3$\\
    $1.24$ & $200$ & $1103$ & $512$ & $12$ & $6144$ & $200$ & $6$ & $1.3236\cdot 10^6$\\
    \hline
    $1.65$ & 200 & $1160$ & $512$ & $12$ & $6144$ & $200$ & $6$ &  $1.392\cdot 10^6$\\
 \hline
 \hline
\end{tabular}
\caption{Number of measurements (last column) for each momentum  (first column) used for computing the disconnected contributions. $N_{\rm ev}$ is the number of eigen-modes (second column), $N_{\rm conf}$ the number of configurations (third column) and $N_{\rm had}$ the number of Hadamard vectors (fourth column). $N_{\rm inv}$ is the number of inversions per configuration (sixth column), computed as the product of the number of stochastic vectors multiplied by $N_{\rm sc}$ (fifth column) which takes into account the spin-color dilution. The number of source positions for the two-point functions $N_{\rm srcs}$ (seventh column)  contributes to the total statistics $N_{\rm meas}$ of the disconnected diagrams, as well as the number of directions and orientations of the nucleon boost $N_{\rm dir}$ (eighth column).}
\label{table:disc_diagrams_stat}
\end{table}

\FloatBarrier
\subsection{Two-point functions}\label{sec:two_point}

The two-point functions enter the calculation through the ratio of Eq.~\eqref{eq:ratio_3pt_2pt}, but also contribute to the evaluation of the disconnected diagram, as shown in Eq.~\eqref{eq:3pt_isoscalar_explicit}. For this reason, to obtain a significant amount of measurements for the disconnected contributions we  compute the two-point functions with a large number of source positions, $N_{\rm src}^{\rm 2pt}=200$, and consider boosts of the  nucleon along the different spatial directions and orientations. This procedure allowed us to considerably reduce the statistical error in the disconnected contributions at small computational cost because the same loops is combined with all 200 two-point functions on the same configurations. Given that the computational cost of the two-points function is considerably lower compared to the one required to evaluate the disconnected quark loops, using multiple source positions is highly beneficial. In Table~\ref{table:disc_diagrams_stat} we report the number of measurements of the two-point function performed at each nucleon boost $P_3$.

\begin{figure}[h!]
    \centering
    \includegraphics[width=\textwidth]{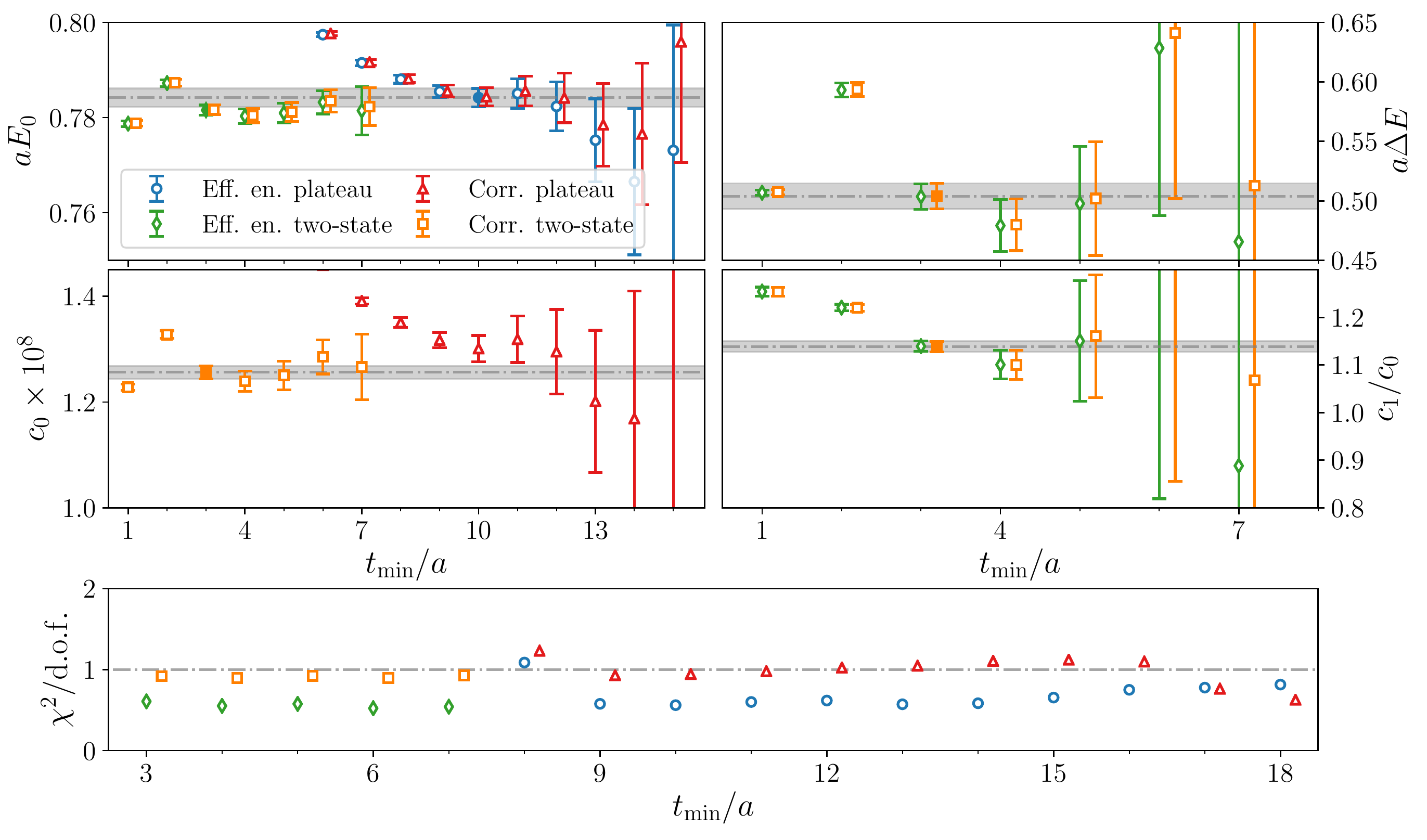}
    \caption{Results of the two-state and plateau fits performed on the two-point correlator and on the effective energy at $P_3=1.24$~GeV as a function of the lowest time, $t_{\rm min}$ used in the fit, using the expansion of  Eqs.~\eqref{eq:two_state_twop} - \eqref{eq:two_state_eff_en}.  In the lower panel we report the reduced $\bar{\chi}^2\def \chi^2/{\rm  d.o.f.}$ for each fitting procedure.  The gray bands correspond with the selected values for $E_0$ and the remaining parameters $\Delta E$, $c_0$, $c_1/c_0$, respectively obtained with the plateau fit of the effective energy and the two-state fit of the correlator. The numerical results for the parameters are reported in Tab.~\ref{table:res_fit_2pt}.}
    \label{fig:sum_fit_p3}
\end{figure}

The two-point function can be written as
\begin{equation}\label{eq:2pt_expansion}
    C_{2pt}(\vec{P};t,0)=\sum\limits_{n}|\braket{\Omega|N(\vec{0},0)|n}|^2e^{-tE_n(P)}\,,
\end{equation}
with $\ket{n}$ being  the $n^{\rm th}$ energy state of the interpolator $N_\alpha(x)$ and $E_n(P)$ its energy. We performed the analysis by keeping up to two terms in the expansion of Eq.~\eqref{eq:2pt_expansion}. In particular, the two-state fit function of the two-point correlator consists of 
\begin{equation}
\begin{split}\label{eq:two_state_twop}
    C_{2pt}(\vec{P};t,0) &= c_0e^{-tE_0}+c_1e^{-tE_1}\\
    &=c_0e^{-tE_0}\left(1+\frac{c_1}{c_0}e^{-\Delta E t}\right)\,,
    \end{split}
\end{equation}
while the effective energy reads
\begin{equation}\label{eq:two_state_eff_en}
    E_{\rm Eff}(\vec{P};t,0)\equiv \log{\left(\frac{C_{2pt}(\vec{P};t,0) }{C_{2pt}(\vec{P};t+1,0) }\right)}=E_0+\log{\left(\frac{1+Be^{-\Delta E t}}{1+Be^{-\Delta E (t+1)}}\right)}\,,
\end{equation}
with $\Delta E=(E_1-E_0)$ and $B=c_1/c_0$. 
In Fig.~\ref{fig:sum_fit_p3}, we show the results of the two-state fits
of the correlator and the effective energy for the parameters $E_0,\,\Delta E ,\,c_0$ and $c_1/c_0$, varying the low-end of the fit interval $t_{\rm min}$. The results show that the fits on the correlator or the effective energy lead to the same ground state energy. Furthermore, the plateau and two-state fits converge at $t_s/a=9$ for momentum 1.24 GeV.

\begin{figure}[h!]
    \centering
    \includegraphics[width=\textwidth]{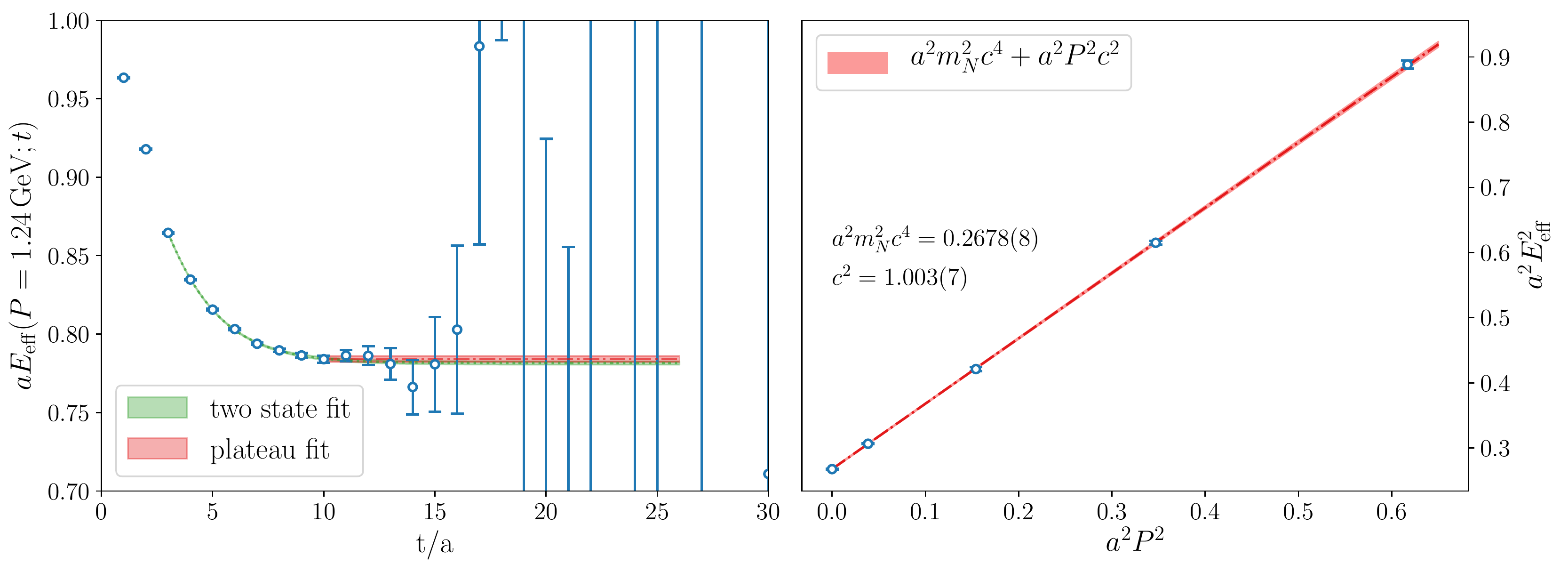}
    \caption{Left panel: effective energy computed for $P_3=1.24\,\gev$, together with the two-state fit (red) and plateau-fit (green) results. Right panel: dispersion relation obtained using the plateau fit for $P_3=0,\,0.41,\,0.83,\,1.24\,\gev$ (blue points). We also report the results for the linear fit using Eq.~\eqref{eq:disp_rel} (red line). }
    \label{fig:disp_relation}
\end{figure}

In Table~\ref{table:res_fit_2pt} we report  the parameters extracted using  one- and two-state fits. The results for $E_0$ are obtained with the plateau fit of the effective energy, and they are compatible with the values extracted using two-state fit results. 
\begin{table}[h]
\begin{tabular}{c | c c c c} 
 \hline
 \hline
 \noalign{\vskip 0.1cm}    
 $P_3\,[\gev]$ & $aE_0$ & $a \Delta E$ & $c_1/c_0$ & $c_0$ \\ [0.5ex]
 \hline\hline
    $0$ & $0.5139(9)$ &  $0.51(9)$ & $0.80(2)$ & $8.99(9)\times 10^{-8}$ \\ 
    $0.41$ & $0.5504(9)$ &  $0.49(2)$ & $0.82(2)$ & $6.74(7)\times 10^{-8}$  \\ 
    $0.83$ & $0.647(4)$ &  $0.48(3)$ & $0.88(4)$ & $3.59(5)\times 10^{-8}$  \\ 
    $1.24$ & $0.784(2)$ &  $0.50(1)$ & $1.14(1)$ & $1.26(1)\times 10^{-8}$ \\
    $1.65$ & $0.942(3)$ &  $0.53(1)$ & $1.34(2)$ & $3.28(5)\times 10^{-9}$ \\
 \hline
 \hline
\end{tabular}
\caption{ Results for the parameters $E_0$, $\Delta E$, $c_1/c_0$ and $c_0$ for $P_3=0,\,0.41,\,0.83,\,1.24,\,1.65\,\gev$. The remaining parameters are obtained with the two-state fit of the two-point correlator of Eq.~\eqref{eq:two_state_twop}.}
\label{table:res_fit_2pt}
\end{table}
In Fig.~\ref{fig:disp_relation} we show the effective energy for the second largest momentum $P_3=1.24\,\gev$, together with the plateau fit and two-state fit results. By iterating the fit procedure described above over the data for the different nucleon boosts, we reconstructed the dispersion relation 
\begin{equation}\label{eq:disp_rel}
   a^2E^2=a^2m_N^2c^4+a^2\vec{P}^2c^2, 
\end{equation}
with $m_N$ being the nucleon mass. In Fig.~\ref{fig:disp_relation} we show the observed trend of the energy with the nucleon boost $P_3$ together with a linear fit performed with the function of Eq.~\eqref{eq:disp_rel}, giving $a^2m_N^2c^4=0.2678(8)$ and $c^2=1.003(7)$. As can be seen, the lattice data are fully compatible with the dispersion relation for all values of $P_3$.

\FloatBarrier
\section{Disconnected matrix elements}\label{sec:disc_mat_el}
\label{sec:disc}

Obtaining the disconnected contributions to the up-, down- and strange-quark PDFs is the central goal, and most laborious aspect of this work. We use the techniques outlined in Section~\ref{sec:disc_contr} to extract the matrix elements and study systematic uncertainties, such as excited-states contamination.

\FloatBarrier
\subsection{Excited-states contamination} \label{sec:excited_states}
To extract reliably the ground-state contribution to the matrix elements, we evaluate the ratio between the three- and two-point functions at seven source-sink separations, ranging from $t_s=0.563\,{\rm fm}$ to $t_s=1.126\,{\rm fm}$ in steps of $a=0.0938\,\fm$. For disconnected contributions, the evaluation of different source-sink separations does not require new inversions. This allowed us to study the excited-states contamination to the matrix elements using several $t_s$ values and three analysis methods: plateau fit, two-state fit and summation method. We briefly summarize these methods.
\begin{enumerate}
\item  {\bf Two-state fit}. In Sec.~\ref{sec:two_point}, the two-point correlator is expanded up to the first excited state. Likewise, we can expand the three-point correlator keeping terms up to the first excited state. This gives four terms, that is
\begin{equation}\label{eq:twost_3pt_fun}
\begin{split}
    C_{3pt}(\vec{P};t_s,\tau) &= \mathcal{A}_{0,0}(\vec{P})e^{-E_0(\vec{P})t_s}\\
    &+\mathcal{A}_{0,1}(\vec{P})e^{-E_0(\vec{P})t_s}e^{-\Delta E (\vec{P})\tau}\\
    & +\mathcal{A}_{1,0}(\vec{P})e^{-E_1(\vec{P}) t_s}e^{\Delta E (\vec{P})\tau}\\
    & +\mathcal{A}_{1,1}(\vec{P}) e^{-E_1(\vec{P})t_s}\,. 
\end{split}
\end{equation}
Being interested in the forward kinematic limit allows one to reduce the number of independent parameters, since $\mathcal{A}_{0,1}=\mathcal{A}_{1,0}$. We performed a fit  of the ratio of Eq.~\eqref{eq:ratio_3pt_2pt} with the function
\begin{equation}\label{eq:twost_thrp_fit_fun}
    \frac{\braket{C_{3pt}(\vec{P};t;t_s,0}}{\braket{C_{2pt}(\vec{P};t,0)}}=\frac{\mathcal{A}_{0,0}}{c_0}\frac{\left[1+\left(\mathcal{A}_{0,1}/\mathcal{A}_{0,0}\right)e^{-\Delta E \tau}+\left(\mathcal{A}_{0,1}/\mathcal{A}_{0,0}\right)e^{-\Delta E (t_s-\tau)}+\left(\mathcal{A}_{1,1}/\mathcal{A}_{0,0}\right) e^{-\Delta E t_s}\right]}{\left[1 + \frac{c_1}{c_0}e^{-\Delta E t_s}\right]}\,,
\end{equation}
where the parameters $c_1/c_0$  and $\Delta E$ are determined through the effective energy fit and the results are reported in Table~\ref{table:res_fit_2pt}. Thus, the parameters determined by fitting the ratio of three- and two-point functions are $\mathcal{A}_{0,0}/c_0$, $\mathcal{A}_{0,1}/\mathcal{A}_{0,0}$ and $\mathcal{A}_{1,1}/\mathcal{A}_{0,0}$. $\mathcal{A}_{0,0}/c_0$ corresponds to the matrix element we are interested in. Such fits are weighted by the statistical errors, and therefore the fit is driven by the most accurate data. Since the statistics for the disconnected contributions are independent of th $t_s$ value, we repeat the two-state fits modifying, each time, the starting value of $t_s$ entering the fit ($t_s^{\rm low}$). The results from the two-state fit method, allows us to verify ground-state dominance by comparing the matrix elements from the individual $t_s$ values. 
\\
\item {\bf Plateau fit}. For $0\ll\tau \ll t_s$ and $\Delta E \,t_s\gg 0$, the first term in the ratio of Eq.~\eqref{eq:twost_thrp_fit_fun} dominates. Thus, the matrix elements can be extracted by performing a constant fit on the ratio of Eq.~\eqref{eq:ratio_3pt_2pt} in the region defined $0\ll \tau \ll t_s$, with large enough source-sink separation. We exclude from the fit range three points from left and right, i.e. we evaluate the weighted average in the interval $\tau \in [3,t_s-3]$. While the excited-states contamination decreases with $t_s$, at the same time the statistical uncertainty exponentially increases. For this reason, the determination of the ground state of the matrix elements with the plateau fit method is a challenging task, and the results need to be compared with other analysis techniques. \\
\item {\bf Summation method}. Summing over the insertion time $\tau$ of the ratio of the three- and two-point functions we find~\cite{Maiani:1987by,Capitani:2012gj}
\begin{equation}
    S(t_s) = \sum\limits_{\tau=2}^{\tau=t_s-2}\frac{\braket{C_{3pt}(\vec{P};t;t_s,0}}{\braket{C_{2pt}(\vec{P};t,0)}} = t_s\frac{A_{0,0}}{c_0}+c + \mathcal{O}\left(e^{-\Delta E t_s}\right).
\end{equation}
Thus, the matrix elements corresponds with the slope of the straight line $S(t_s)$, and can be measured by performing a linear regression.  

\end{enumerate}

Using the three aforementioned approaches we analyze  the excited-states effects on the matrix elements for the unpolarized, helicity and transversity PDFs. In the next three subsections, we present the analysis of the real and imaginary parts of the matrix elements at $P_3=1.24\,\gev$, as a representative example.

\FloatBarrier
\subsubsection{Unpolarized}
\label{sec:unpol_excited}

\begin{figure}[h!]
    \centering
    \includegraphics[scale=0.6]{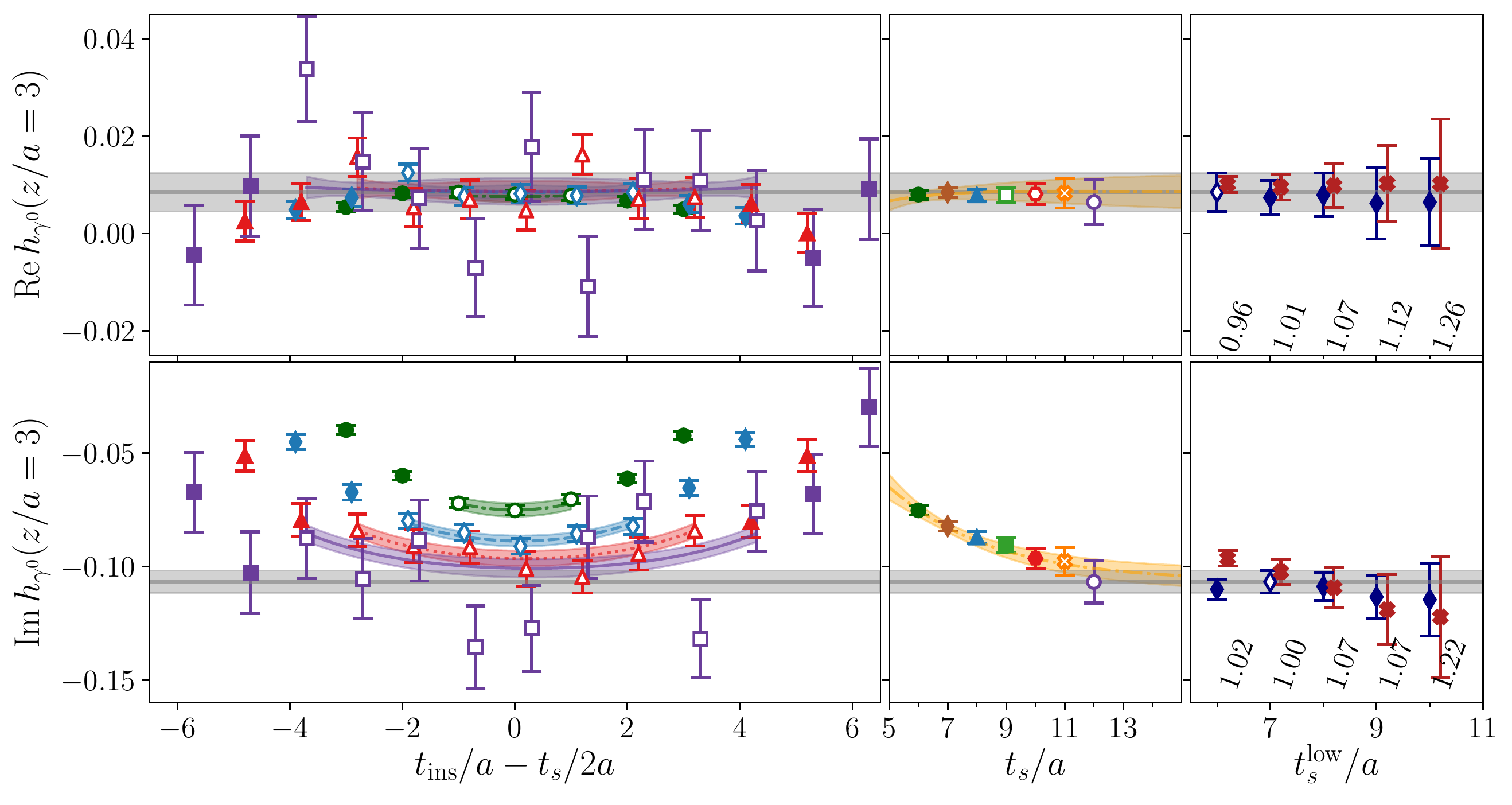}
    \caption{ Left: Results on $C_{3pt}(t;t_s)/C_{2pt}(t)$ for the unpolarized PDFs for  $P_3=1.24\,\gev$, at $t_s/a=6,8,10,12$ for $z/a=3$. The data for $t_s/a=7,9,11$ are omitted to improve the readability. The two-state fit results (gray band), and the value of the two-state fit of Eq.~\eqref{eq:twost_3pt_fun} evaluated at the same $t_s$ as the data-points are also shown. Only the data-points with open symbols are taken into account in the two-state fit procedure. Center: the plateau fit results as a function of $t_s/a$. Each source-sink separation is associated with a different color. The orange band is the predicted $t_s$ dependence of the function in Eq.~\eqref{eq:twost_3pt_fun} at $t_{\rm ins}=t_s/2$. Our final value for the matrix elements is determined as the correlated constant fit of the plateau values shown with open symbols. Right: results of the two-state fit (navy blue) as a function of the lowest source-sink separation $t_s^{\rm low}$ included in the fit. The empty data-point is the selected two-state fit result, which corresponds to the gray band. For each $t_s^{\rm low}$ we report the reduced $\chi^2$ of the two-state fit. The results obtained with the summation method are reported with the red open crosses as a function of $t_s^{\rm low}$.}
    \label{fig:summary_z3_exc_st_unp}
\end{figure}

We start by discussing the analysis of the unpolarized isoscalar $u+d$ disconnected matrix elements. In Fig.~\ref{fig:summary_z3_exc_st_unp} we show the ratio of three- and two-point functions at $P_3=1.24\,\gev$ for the unpolarized operator for $z/a=3$ and we compare the results obtained with the three analysis methods reported in Sec.~\ref{sec:excited_states}. The real part of the matrix elements shows no substantial dependence on the source-sink separation, and the plateau fit results obtained at different $t_s$ give all compatible results. The dependence of the two-state fit on the lowest source-sink separation $t_s^{\rm low}$, included in the fit shows a constant trend, which is also compatible with the results obtained with the summation method. The reduced chi-square $\chi^2/{\rm d.o.f}=0.96$  suggests that the function of Eq.~\eqref{eq:twost_thrp_fit_fun} provides a good description of the data. To extract the matrix elements, we compute the constant correlated fit of the plateau fit results starting from $t_s^{\rm low}/a=9$.
In contrast to the real part, the imaginary part shows a large effect due to the excited-states contamination. However, the two-state fit is compatible with the plateau value using $t_s/a=11$. Also, the results obtained at different $t_s^{\rm low}$ using the summation method are compatible with the other methods within uncertainties. As final results for the unpolarized matrix element, we report the ones from the plateau fit for $t_s/a=11$,  which is compatible with the results obtained with the two-state fit and summation method.

\FloatBarrier
\subsubsection{Helicity}\label{sec:ex_stat_helicity}
The disconnected contributions to the helicity isoscalar matrix elements is purely real and exhibits a non-negligible dependence on the source-sink separation. We observe a decreasing behavior as $t_s$ increases, which is a behavior also observed in the axial charge~\cite{Alexandrou:2019brg}. In addition, the plateau fit for $t_s/a> 10$ are compatible with the two-state fit. Therefore, we use the plateau fit for $t_s/a=10$ as our final results, so that statistical uncertainties are controlled. 

\begin{figure}[h!]
    \centering
    \includegraphics[scale=0.6]{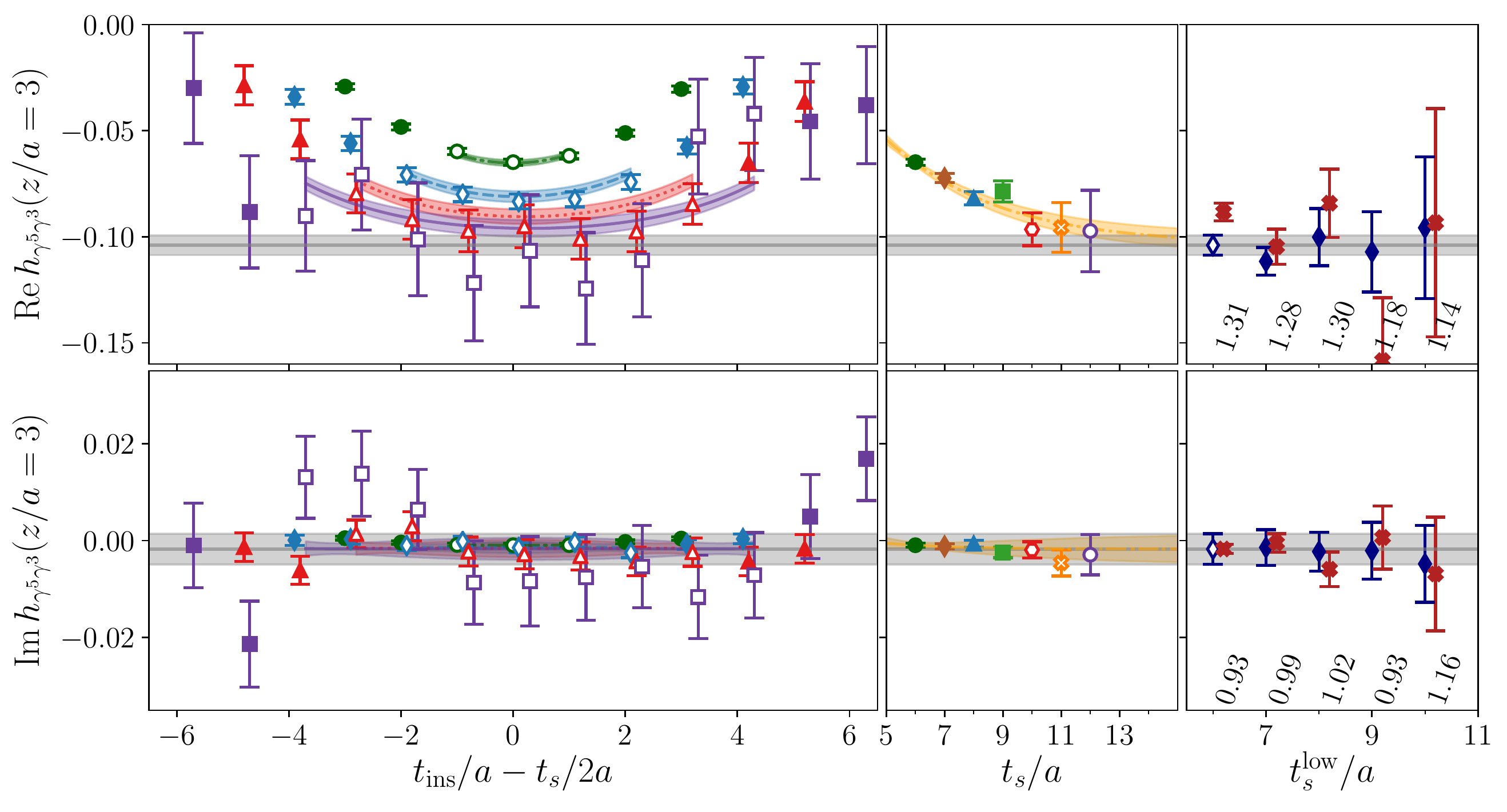}
    \caption{The same as Fig.~\ref{fig:summary_z3_exc_st_unp} but for  the isoscalar helicity matrix elements.}
    \label{fig:summary_z3_exc_st_hel}
\end{figure}

\FloatBarrier
\subsubsection{Transversity}
\label{sec:transv_excited}

\begin{figure}[ht!]
    \centering
    \includegraphics[scale=0.6]{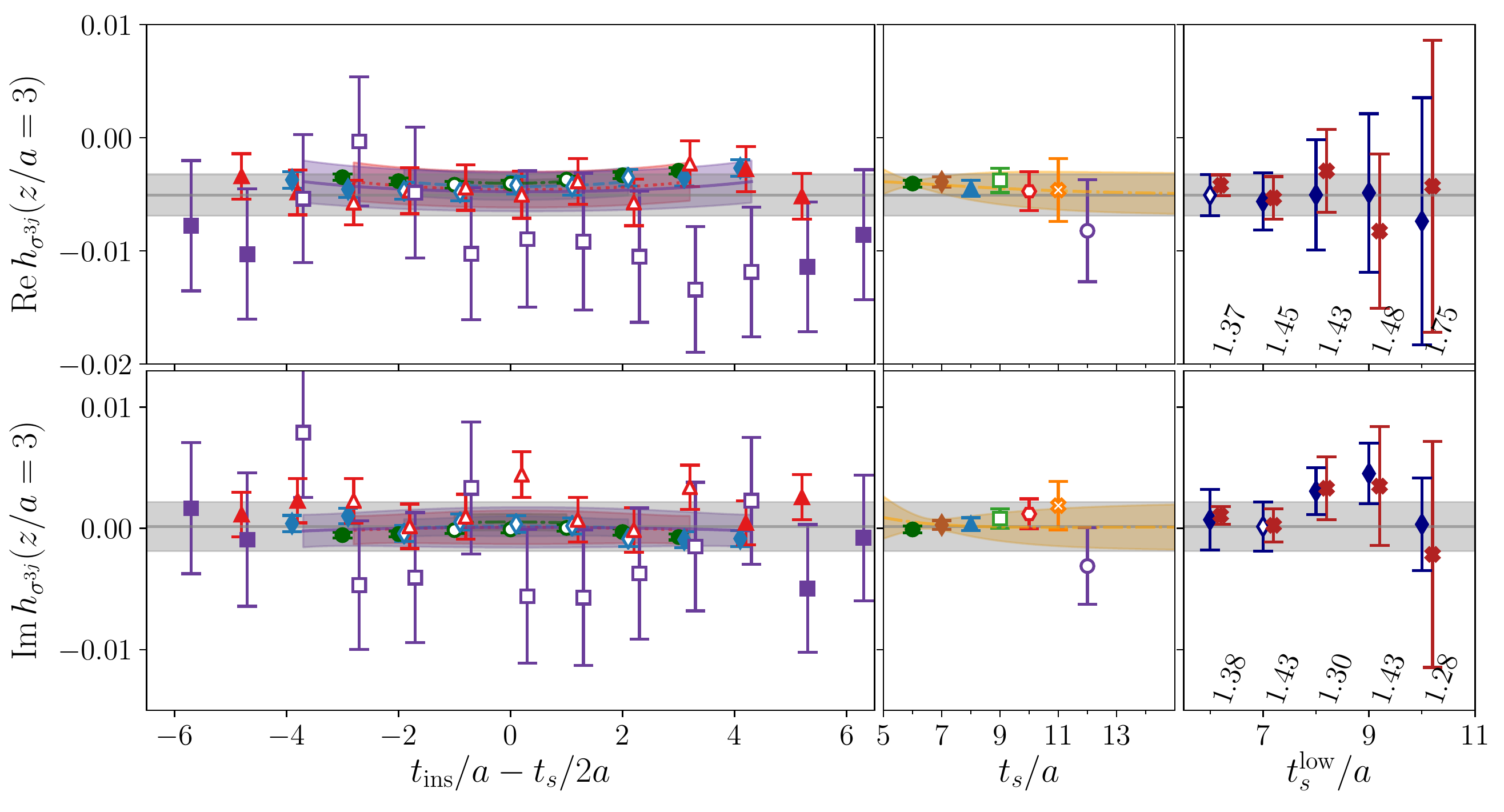}
    \caption{{The same as Fig.~\ref{fig:summary_z3_exc_st_unp} but for  the isoscalar transversity matrix elements. }}
    \label{fig:summary_z3_exc_st_tr}
\end{figure}

\begin{figure}[h!]
    \centering
    \includegraphics[width=\linewidth]{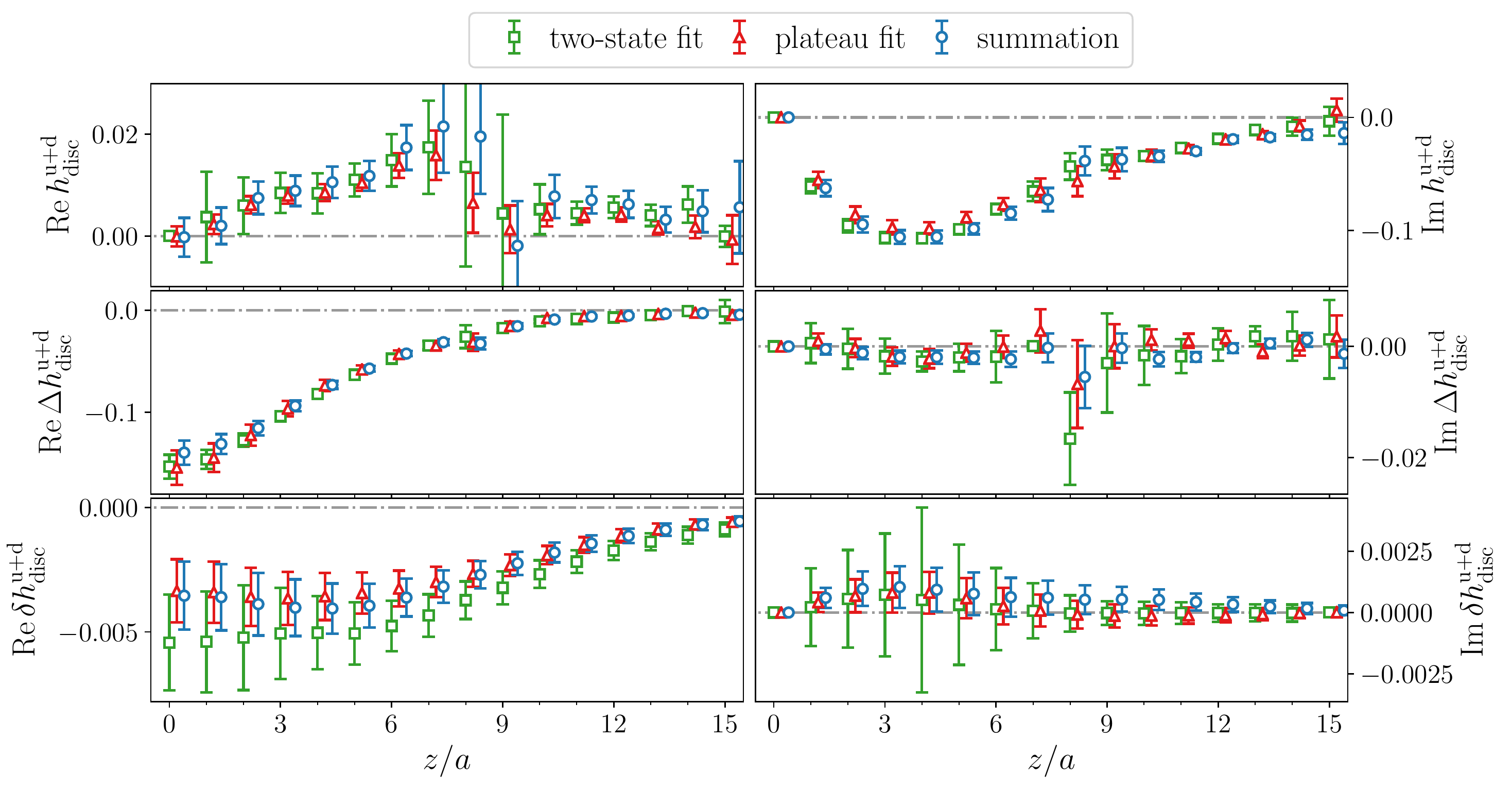}
    \caption{Comparison of the matrix elements obtained from the one- (red points) and two-state (green points) fits and the summation method (blue points) for the disconnected isoscalar matrix elements at $P_3=1.24\,\gev$. From top to bottom we show the unpolarized, helicity and transversity PDFs. See text for more details.}
    \label{fig:comparison_m3_diff_meths}
\end{figure}

The ratio of the three- and two-point functions for the disconnected contributions to the isoscalar transversity matrix elements does not shows dependence on the source-sink separation for both the real and imaginary parts (see Fig.~\ref{fig:summary_z3_exc_st_tr}). Thus, the matrix elements are computed from the plateau fit for $t_s/a=9$, both for the real and the imaginary parts. 
\FloatBarrier
Using the criterion adopted for selecting the final results for each PDF case we compared the extracted matrix elements in Fig.~\ref{fig:comparison_m3_diff_meths} for all values of $z$. In summary, the plateau fits are evaluated at $t_s/a= 9,\,10,\,9$ ($t_s/a=11,\,10,\,9$) for the real (imaginary) part of the unpolarized, helicity and transversity, respectively.  For the summation method, 
we employ all $t_s$ values available, except for the imaginary part of the unpolarized and the real part of the helicity matrix elements, where $t_s/a=6$ is excluded as explained above. The two-state fit is performed in the range $t_s/a \in[6,12]$ in all cases, except for the imaginary part of the unpolarized and transversity matrix elements, where the $t_s/a=6$ value, as explained, is not included in the regression.

Our conclusions for the isoscalar disconnected matrix elements apply also to the strange matrix elements, with the excited-states contamination showing similar effects. 

\FloatBarrier
\subsection{Momentum dependence}\label{sec:mom_dep_disc}

\begin{figure}[ht!]
    \centering
    \includegraphics[width=\linewidth]{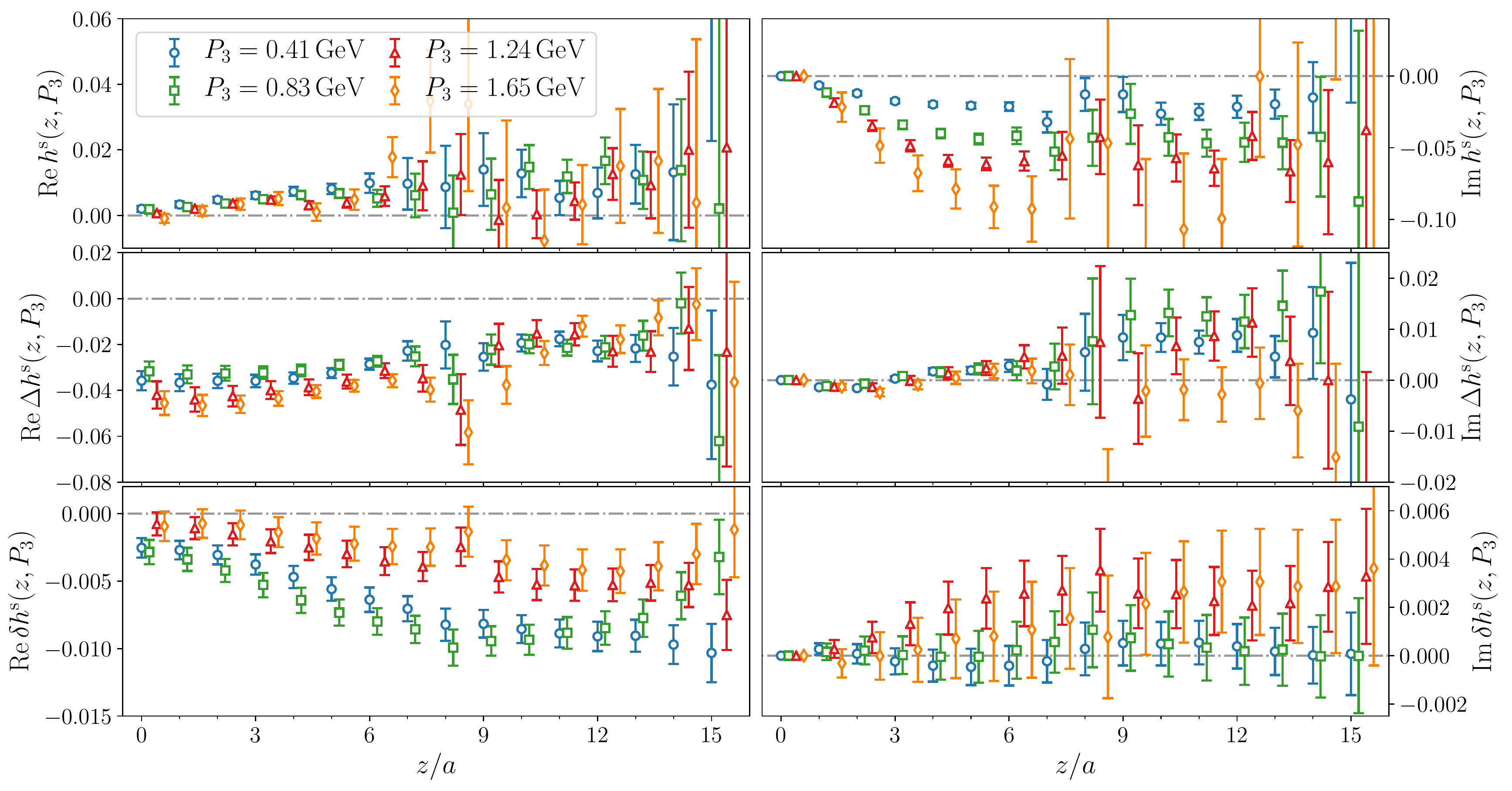}
    \includegraphics[width=\linewidth]{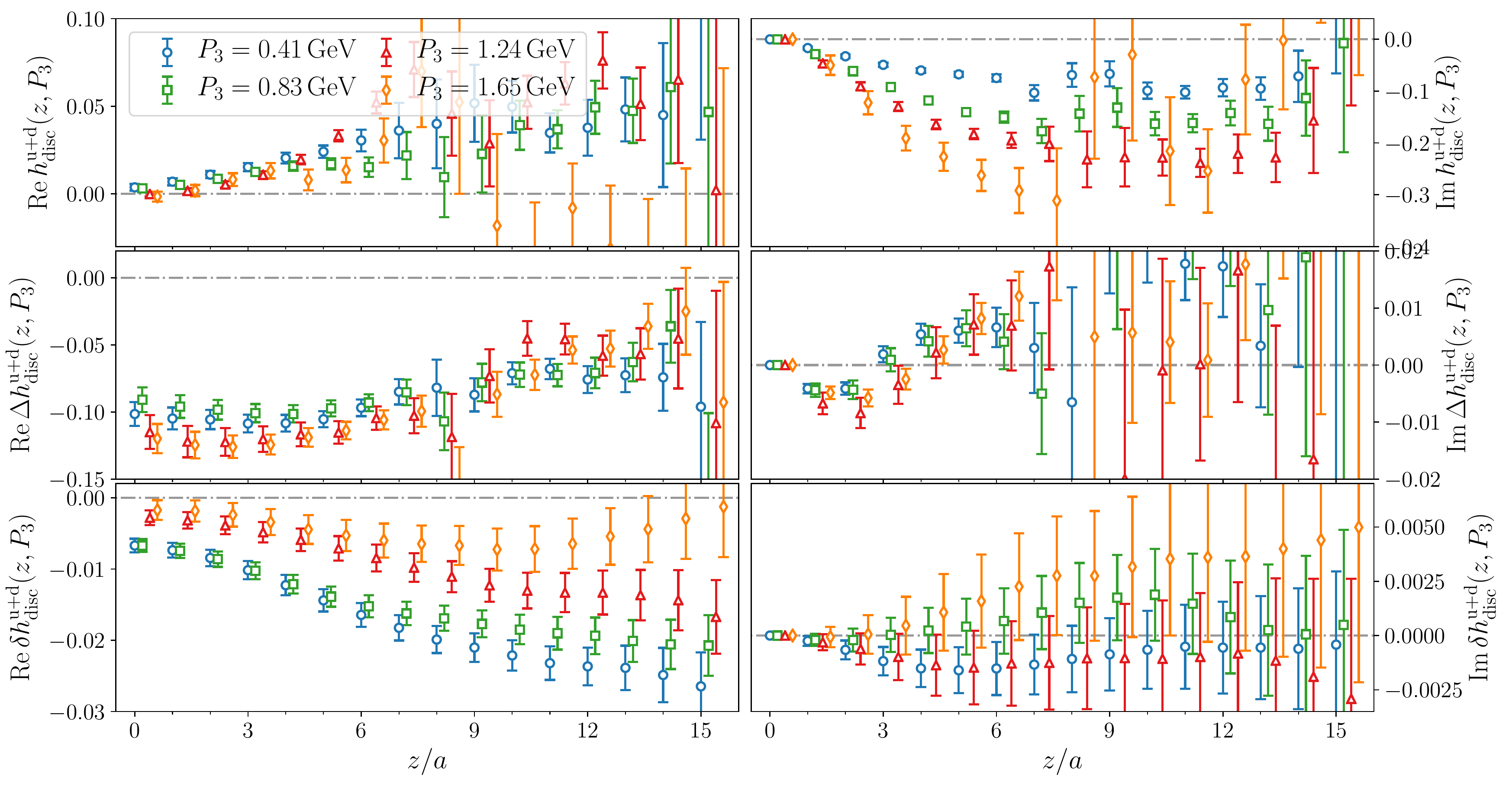}
    \caption{Momentum dependence of the renormalized matrix elements for the strange (upper figure) and isoscalar disconnected (lower figure) unpolarized (top panels), helicity (middle panels) and transversity (bottom panels) distributions. We show the matrix elements computed at $P_3=0.41\,\gev$ (blue), $0.83\,\gev$ (green), $1.24\,\gev$ (red) and $1.65\,\gev$ (yellow). Data points are slightly shifted to improve readability.}
    \label{fig:mom_dep_str_me}
\end{figure}

As explained in Sec.~\ref{sec:quasiPDF}, the matrix elements and the quasi-PDFs have a dependence on the nucleon boost $P_3$, which also enters the matching formula leading to the PDFs. An important aspect of the study is the investigation of the momentum dependence of the matrix elements, which affects the convergence to the light-cone PDFs. In Fig.~\ref{fig:mom_dep_str_me}, we present the results for the renormalized strange and isoscalar disconnected matrix elements as a function of the momentum boost. For the unpolarized case, the real part decreases in magnitude as the $P_3$ increases, and becomes compatible with zero. In contrast, its imaginary part is non-zero and shows convergence for the two largest values of $P_3$. We find that the isoscalar disconnected matrix elements share the same qualitative behavior as the strange-quark ones.

First results on the helicity distribution appeared in Ref.~\cite{Alexandrou:2020uyt}. Here we show results  with increased statistics, and with the addition of $P_3=1.65\,\gev$. The matrix elements show a mild residual dependence on the momentum. The imaginary part of the renormalized matrix elements arises entirely from the complex multiplication with the renormalization function and the bare matrix elements (see Eq.~(\ref{eq:renorm})). Indeed, as  mentioned already in Sec.~\ref{sec:ex_stat_helicity}, the disconnected contribution to the bare matrix element for the helicity distribution is purely real. 

The real part of the matrix elements for the transversity distribution exhibit a strong dependence on the nucleon boost, changing dramatically as we increase $P_3$ from $0.83\,\gev$ to $1.24\,\gev$. However, results obtained for  $P_3=1.65\,\gev$ show agreement with those for  $P_3=1.24\,\gev$  albeit the large uncertainty. From the current results it is still unclear if convergence is reached. However, in order to fully check this would require a larger momentum and much more measurements to reach the required accuracy. This is beyond the current study and will be tested in a followup work. We thus, construct the PDFs using the results for $P_3=1.24$~GeV. In Sec.~\ref{sec:strange_distr}, we comment on the region of $x$ affected by the gap observed in the real part of the matrix elements as momentum increases. In contrast, the imaginary part is fully compatible with zero for the two lowest momenta, while it is slightly non-zero at large $z$ at the highest momentum. 

\FloatBarrier
\section{Connected matrix elements}\label{sec:conn_mat_el}

\begin{figure}[h!]
    \centering
    \includegraphics[width=\linewidth]{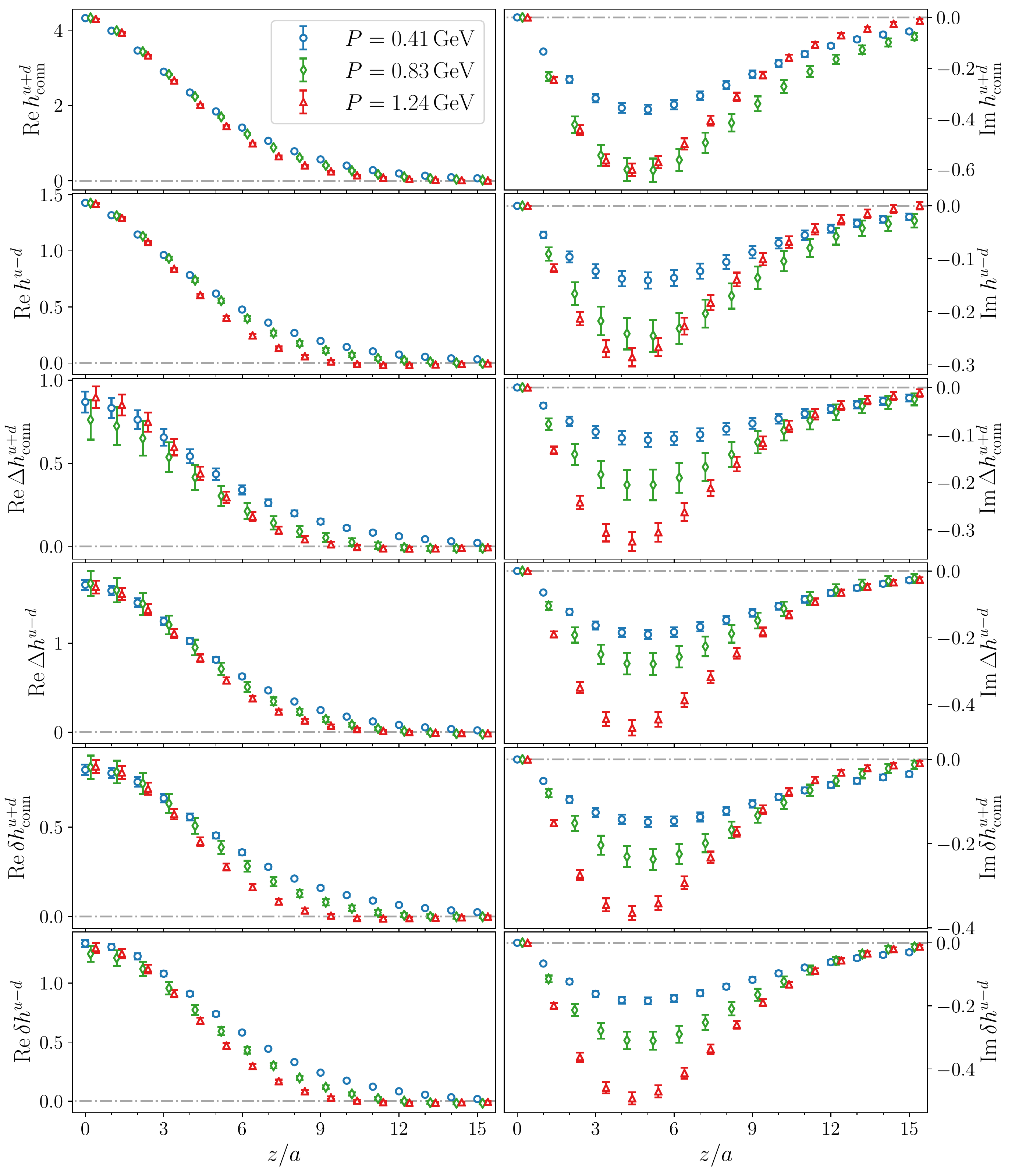}
    \caption{Momentum dependence of the bare connected contributions to the matrix elements. The first two rows show respectively the isoscalar  connected contribution and the isovector  unpolarized matrix elements. The left column shows the real part and the right the imaginary part. The same flavor combinations are reported respectively for the helicity and transversity distributions  in the 3$^{\rm rd}$ and 4$^{\rm th}$ rows, and in the last two rows. We show the matrix elements at $P_3=0.41\,\gev$ (blue), $0.83\,\gev$ (green) and $1.24\,\gev$ (red). Data points are slightly shifted to to improve readability.  }
    \label{fig:bare_conn_me}
\end{figure}

The evaluation of the connected matrix elements contributing to the three types of PDFs has been studied in our previous works. In particular, we refer the reader to the study of Ref.~\cite{Alexandrou:2019lfo}, where several sources of systematic uncertainties were discussed in great detail. For completeness, we briefly discuss here the connected contributions which are needed for the flavor decomposition. In Fig.~\ref{fig:bare_conn_me}, we show the momentum dependence of the bare connected contributions to the isoscalar and isovector matrix elements for the three types of PDFs. In all cases, as the nucleon boost increases, the real part of the matrix elements decay to zero faster, and the magnitude  for the imaginary part increases in the region $z/a\lesssim 9$.  The unpolarized matrix elements show convergence with $P_3$ while the imaginary parts of  the helicity and transversity distributions increase in magnitude. We note that in order to compute the connected contributions for a fourth larger boost would require new inversions and large number of measurements to reduce the errors sufficiently enough to check convergence. Thus for the current work we opt to use the results for $P_3=1.24$~GeV for the connected parts since the focus of this work is the evaluation of the disconnected contributions. The uncertainty on the unpolarized distribution is smaller as compared to the other two distributions. This behavior is due to the fact that both the three- and two-point functions share the same projector, $(1+\gamma^0)/2$, which increases the correlation between the two quantities and, as a result, drastically decreases the noise-to-signal ratio.   We note also that the transversity distribution reported here is the average over the two insertions $\sigma_{3j}$ with $j=1,2$. 

\FloatBarrier
\section{Nucleon charges}
\label{sec:charges}
The nucleon charges are usually extracted from the nucleon matrix elements of local operators. This limit is obtained from the matrix elements of non-local operators at $z=0$. Since the charges are frame independent, any value of $P_3$ may be used. Indeed, in Sec.~\ref{sec:conn_mat_el} we demonstrate that the $z=0$ have little dependence on $P_3$  For the disconnected contributions, we have the matrix elements at $P_3=0$ (rest frame). For the connected contributions to the charges, we use the lowest momentum, so we control statistical uncertainties.  In what follows, we will show the results obtained for the isovector $u-d$,  isoscalar $u+d$ and strange-quark vector, axial and tensor charges, $g_V$,\,$g_A$ and $g_T$. 

\begin{figure}[h!]
    \centering
    \includegraphics[scale=0.55]{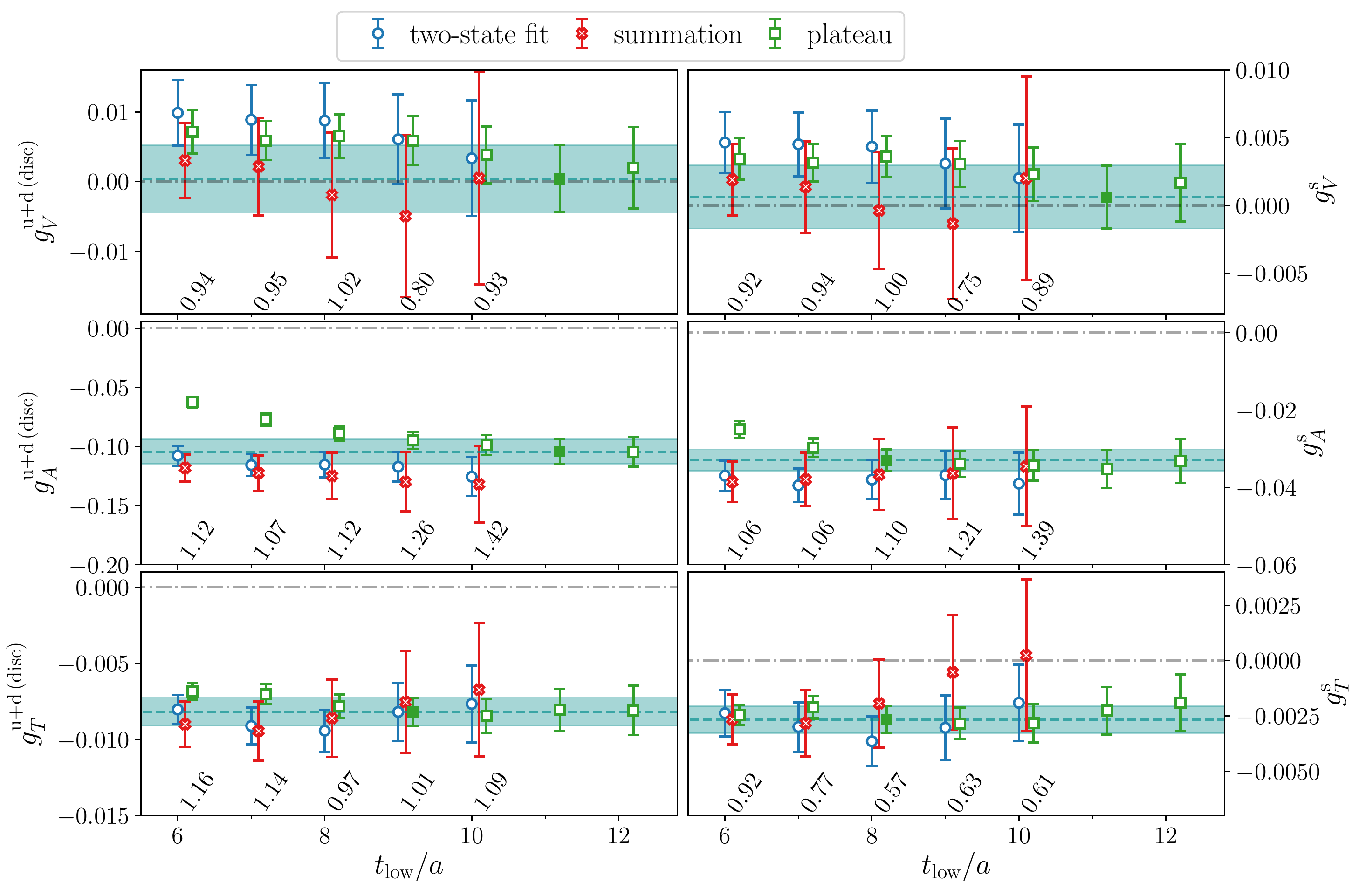}
    \caption{Isoscalar $u-d$ (left) and strange (right) disconnected contributions to the renormalized $g_V$ (top panels), $g_A$ (middle panels) and $g_T$ (bottom panels). In each subplot we show the results obtained with the plateau fit (open green squares), two-state fit (open blue circles) and summation method (open red crosses) as a function of $t_{\rm low}$. We also include $\chi^2/{\rm d.o.f.}$ for the two-state fit. The horizontal band corresponds to the selected plateau fit result. }
    \label{fig:charges}
\end{figure}
First, we describe our results for the disconnected contributions, whose total number of measurements in the rest frame is $N_{\rm meas}=66\cdot 10^3$.
In Fig.~\ref{fig:charges} we show our results for the renormalized disconnected vector, axial and tensor charges. The integral over the volume (i.e. Fourier transform at zero momentum transfer) of the trace of the vector current $\bar{\psi}(x)\gamma^0\psi(x)$ is zero because quark and antiquark loops contributions cancel each other. Thus, the disconnected contribution to the unpolarized isoscalar and strange matrix elements in the absence of the Wilson line and at $P_3=0$ expected to be zero is verified and this  constitutes a consistency  check  of our computations. Due to excited-states contamination, the disconnected isoscalar and strange vector charges $g_V^{\rm u+d\,(disc)}$ and $g_V^{s}$ obtained with the two-state and plateau fits are not compatible with zero at small source-sink separation. In particular, the two-state fit results become compatible with zero when $t_{\rm low}\geq10$. The results obtained with the summation method have the largest uncertainties and are  compatible with zero. \\
The axial charge shows the larger contamination from excited states. In particular, the plateau fit results show a decreasing trend with the $t_{\rm low}$, converging to a constant value for $t_{\rm low}/a\geq11$ for the isoscalar and $t_{\rm low}/a\geq8$ for the strange charges, which are selected as our final values. In contrast, both the results obtained with two-state fit and summation are constant  and  compatible with the selected plateau fit results. We find 
\begin{equation}
    g_A^{\rm u+d\,(disc)}=-0.104(10),\quad g_A^{s}=-0.0320(28)\,.
\end{equation}

The results on the disconnected contributions of the tensor charge  show very mild  excited states effects. We use the plateau value extracted by fitting the ratio to a constant for $t_s/a\geq 9$ for the isoscalar connected tensor charge $g_T^{\rm u+d(disc)}$ and for $t_s/a\geq 8$ for $g_T^s$. Our final results for these two quantities are
\begin{equation}
    g_T^{\rm u+d\,(disc)}=-0.00818(91),\quad g_T^{s}=-0.00265(60)\,.
\end{equation}
We stress that despite the agreement of the strange tensor charge with the value extracted using local operators~\cite{Gupta:2018lvp,Alexandrou:2017qyt}, a direct comparison is not meaningful since we are using gauge ensembles simulated with  heavier than physical pion mass. It thus comes with no surprise that the disconnected isoscalar tensor charge differs from the value obtained at the physical pion mass. 


The connected contributions to the nuclear charges are computed for the smallest momentum  $P_3=0.41\,\gev$. Using these results we can extract the values for each quark flavor for the vector, axial an tensor charges. The details on the computation of the connected isoscalar and isovector contributions are given in Sec.~\ref{sec:conn_mat_el}.
The connected contributions used to extract the charges are obtained from plateau fits with $t_s/a=12$. The nucleon axial and tensor charges are given in Table~\ref{tab:res_charges}. We note that for the vector charge $g_V$ we find results that are consistent with charge conservation.
\begin{table}[h]
\begin{tabular}{c | c c c c c c} 
 \hline
 \hline
\noalign{\vskip 0.1cm}    
   &$u-d$ & $u+d$ (conn.) & $u+d$ (disc.) & $u$ & $d$ & $s$ \\[0.5ex]
 \hline\hline
    $g_A$ & 1.25(4) &  0.66(7) & -0.104(10) & 0.90(2) & -0.35(2) & -0.0320(28)\\ 
    $g_T$ & 1.11(2) &  0.68(2) & -0.00818(91) & 0.89(1) & -0.22(1) & -0.00265(60)\\
 \hline
 \hline
\end{tabular}
\caption{ Results for the isovector (first column), isoscalar connected (second column) and disconnected (third column) and for the up (fourth column), down (fifth column) and strange (sixth column).  We show our results on  the axial (second row) and tensor (third row) nucleon charges.  }
\label{tab:res_charges}
\end{table}

\FloatBarrier
\section{Parton distribution functions}
\label{sec:PDFs}
\subsection{Isoscalar and isovector renormalized matrix elements}\label{sec:ren_conn_me}

In Fig.~\ref{fig:ren_me} we show the momentum dependence of the total renormalized isoscalar and isovector matrix elements, including disconnected contributions. The renormalized matrix elements are reported as a function of $zP_3$. We renormalize and apply the matching procedure independently for the isoscalar and isovector distributions, allowing us to obtain the individual up and down  quark PDFs. 

The source-sink separation used is $t_s=0.94$~fm for the lowest momentum and $t_s=1.13\,\fm $ for $P_3=0.83$ and $1.24\,\gev$. From previous studies of the isovector distributions (see, e.g. Ref.~\cite{Alexandrou:2019lfo}) and nucleon charges~\cite{Alexandrou:2019brg}, we expect that excited-states contamination is more significant for the nucleon three-point correlators of the axial and tensor currents as compared to the vector. However, the statistical uncertainty is larger for these quantities and within our current errors the source-sink separation employed at the highest momentum is  sufficient  to suppress excited states to this level of accuracy~\cite{Alexandrou:2019lfo}.

A summary of the results for the connected matrix elements in the absence of the Wilson line are reported in Table~\ref{tab:mat_el_z0}. The momentum dependence of all the matrix elements analyzed is negligible for $z=0$ as expected 
For example, the isoscalar connected matrix elements for the unpolarized distribution, $h^{\rm u+d}(z=0)$, for the largest momentum differs from the others by less then $1\%$. The isovector unpolarized matrix elements at $z=0$ is independent of the momentum boost, and equal to 1, as expected from charge conservation.
Regarding the isoscalar helicity case, we still find agreement for different $P_3$ within uncertainties, but with larger fluctuations of the mean values, as the disconnected contribution is about $\sim 17\%$ of the connected part. We note that the $\Delta h^{u-d}(z=0)$ is compatible with the experimental value $g_A^{\rm u-d}=1.27641(56)$~\cite{Markisch:2018ndu}. Insensitivity to the momentum boost is also observed in the transversity case.

\begin{table}
\setlength{\tabcolsep}{5pt}
\begin{tabular}{c | c c | c c | c c} 
 \hline
 \hline
    $P_3$ & $h^{u-d}(z=0)$ & $h^{u+d}(z=0)$ & $\Delta h^{u-d}(z=0)$ & $\Delta h^{u+d}(z=0)$ & $\delta h^{u-d}(z=0)$ & $\delta h^{u+d}(z=0)$  \\[0.5ex]
 \hline\hline
    $0.41\,\gev$ & 1.005(4) &  3.046(4) & 1.25(4) & 0.52(5) & 1.11(2) & 0.67(2) \\ 
    $0.83\,\gev$ & 1.004(8) &  3.053(8) & 1.26(11) & 0.45(9) & 1.04(6) & 0.69(5)\\ 
    $1.24\,\gev$ & 1.000(4) &  3.026(5) & 1.23(5) & 0.52(5) & 1.08(3) & 0.69(3)\\
 \hline
 \hline
\end{tabular}
\caption{ Momentum dependence of the unpolarized, helicity and transversity isovector (first column) and isoscalar (second column) matrix elements at $z=0$.  }
\label{tab:mat_el_z0}
\end{table}

\FloatBarrier
\subsection{Truncation of the Fourier transform}\label{sec:syst_eff_truncation}

In order to construct the x-dependence of PDFs we need to take the Fourier transform. Since the matrix elements are determined for discrete finite number of $z$ values, we study the dependence on the cutoff $z_{\rm max}$ to understand systematic effects related to the reconstruction. In particular, for all types of distributions we verify that a $z_{\rm max}$ exists such that, addition of information for $z > z_{\rm max}$ in the Fourier transform leaves the PDF unchanged within statistical uncertainties. This value of $z$ is selected as the maximum value $z_{\rm max}$ included in the Fourier transform and, typically, the matrix elements at this value has a vanishing real part. Note that the latter is just a qualitative criterion and in practice we always check by increasing $z$ for convergence. 

\begin{figure}[h!]
    \centering
    \includegraphics[width=0.9\linewidth]{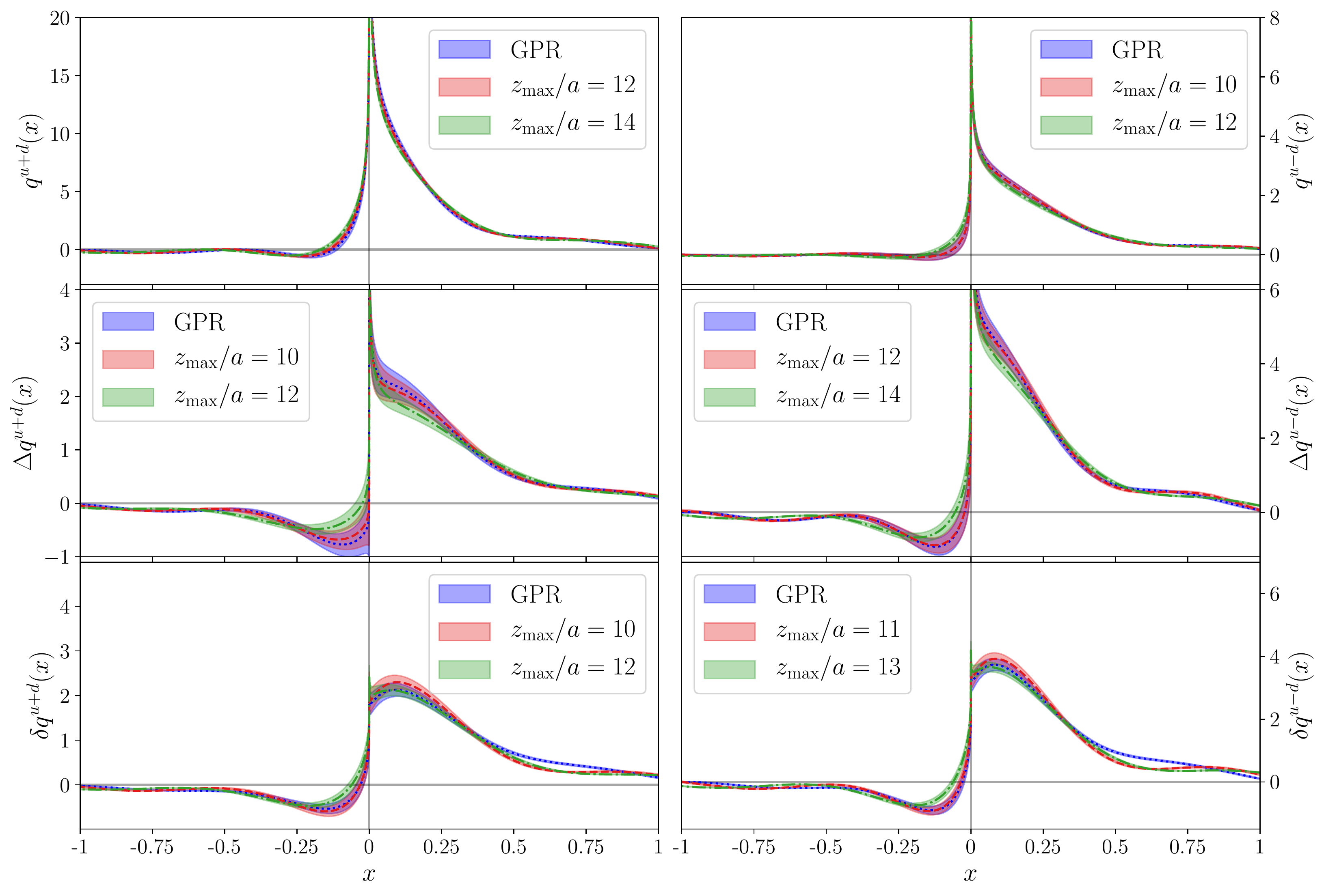}
    \vspace*{-0.2cm}
    \caption{Cutoff dependence ($z_{\rm max}$) of the isoscalar (left) and isovector (right) unpolarized (upper panels), helicity (middle panels) and transversity (bottom panels) at $P_3=1.24\,{\rm GeV}$. Results from BGFT are shown with a blue band. The distributions corresponding to the value of the cutoff reported in Tab.~\ref{table:cutoff_table} are reported in red.}
    \label{fig:zmax_dep_vecsca}
\end{figure}

In Fig.~\ref{fig:zmax_dep_vecsca} we show the dependence on the cutoff $z_{\rm max}$ for the isoscalar and isovector distributions at $P_3=1.24\,{\rm GeV}$. We compare the results obtained with the discrete Fourier transform of Eq.~(\ref{eq:quasi-PDF-discrete}) with the results from the Bayes-Gauss-Fourier transform (BGFT)~\cite{Alexandrou:2020tqq}. The latter is an advanced reconstruction technique based on Gaussian process regression, which allows to obtain an improved estimate of quasi-PDF for continuous values of $x$, starting from a discrete set of data obtained with lattice QCD computations. The chosen values of $z_{\rm max}$ for each quark flavor and each operator are given in Table.~\ref{table:cutoff_table}. From Fig.~\ref{fig:zmax_dep_vecsca} it is clear that increase of the cutoff beyond the reported values of Table~\ref{table:cutoff_table}, does not affect the results for the PDFs. We also find compatible of the discreet Fourier transform with the results using BGFT.


\begin{table}[h!]
\begin{tabular}{c | c c c } 
 \hline
 \hline
 \noalign{\vskip 0.1cm}    
    & Isoscalar & Isovector & Strange\\ 
    \hline\hline
    Unpolarized & 15,14,12 & 15,13,10 & 15,12,12 \\
    Helicity & 15,10,10 & 15,12,12 & 14,14,14\\
    Transversity & 15,11,10 & 15,11,11 & 14,12,12\\
  \noalign{\vskip 0.1cm}  
 \hline
 \hline
\end{tabular}
\caption{Values of $z_{\rm max}/a$ used in the Fourier transform for each type of distribution. Each triplet of numbers corresponds to the cases  for $P_3=0.41,0.83$ and $1.24\,\gev$, respectively.}
\label{table:cutoff_table}
\end{table}

\FloatBarrier
\subsection{Isoscalar and isovector distributions}\label{sec:vec_and_sca_distr}

\begin{figure}[h!]
    \centering
    \includegraphics[scale=0.55]{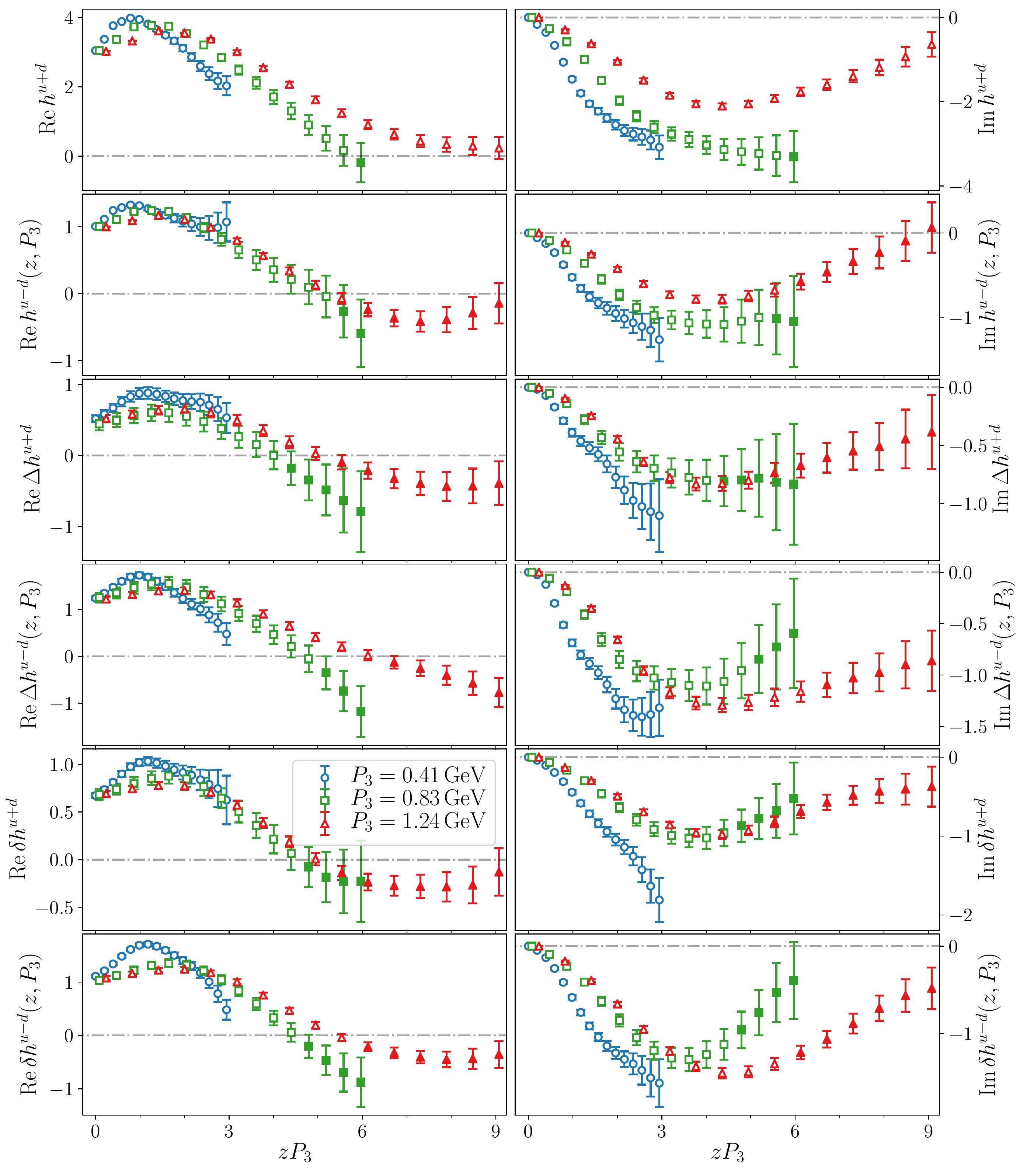}
    \caption{Real (left) and imaginary (right) parts of the renormalized matrix elements as a function of $zP_3$. From top to bottom and in rows of two we show the isoscalar and the isovector matrix elements for the unpolarized, helicity and transversity cases, respectively. The points included in the Fourier transform of Eq.~\eqref{eq:quasi-PDF} are shown with open symbols. Each sub-figure shows the momentum dependence of the corresponding matrix element, where the blue circles correspond to $P_3=0.41\,{\rm GeV}$, the green squares to $P_3=0.83\,{\rm GeV}$ and the red triangles to $P_3=1.24\,{\rm GeV}$.
    }
    \label{fig:ren_me}
\end{figure}

The isoscalar and isovector  PDFs are extracted from the corresponding renormalized matrix elements shown in Fig.~\ref{fig:ren_me}. For the isoscalar combination, we add both the connected and disconnected contributions. We plot the matrix elements against $zP_3$, which is the argument of the exponential in the Fourier transform.

\begin{figure}[h!]
    \centering
    \includegraphics[width=0.9\linewidth]{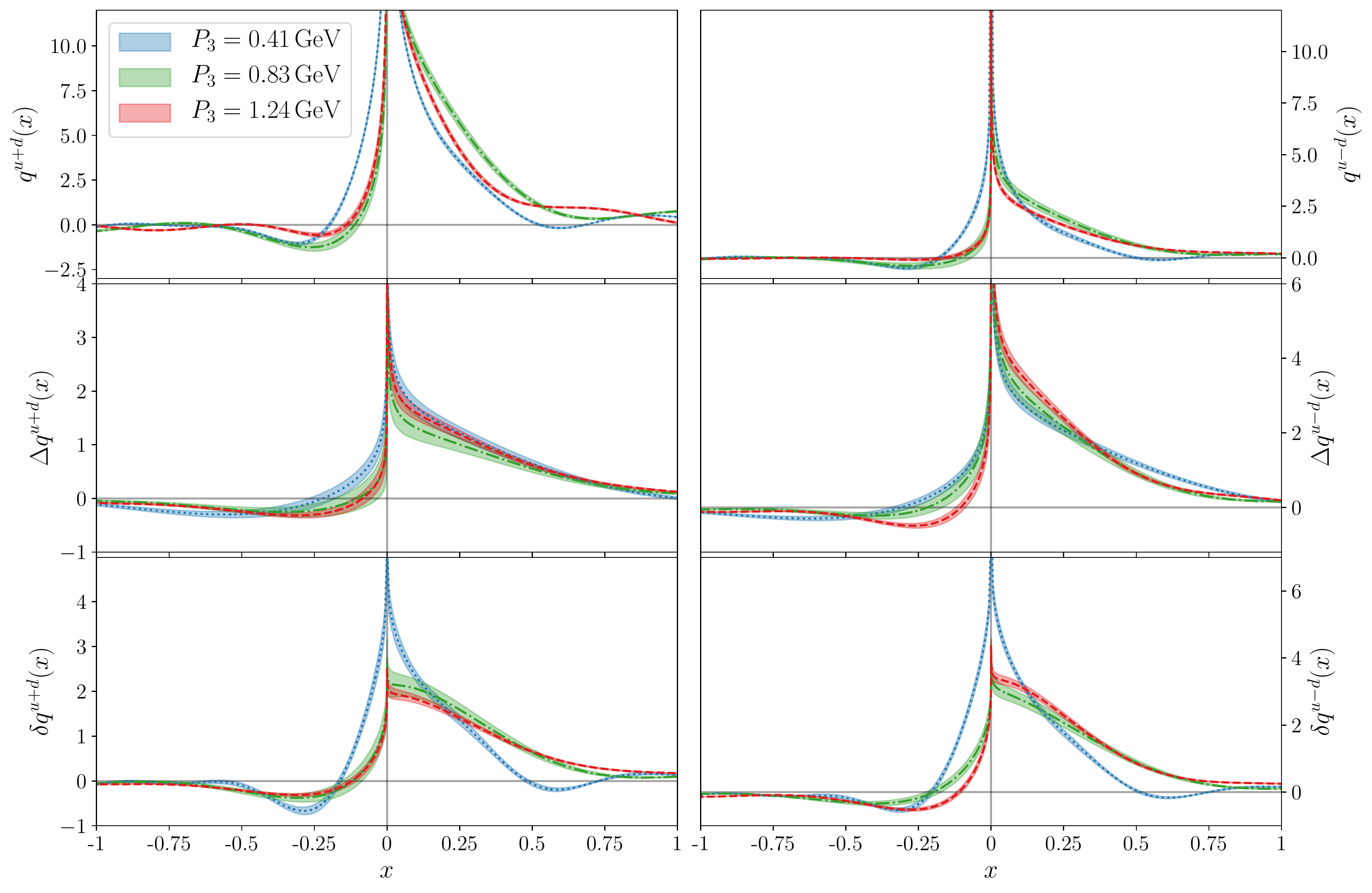}
    \caption{Results for the isoscalar (left) and isovector (right) unpolarized (first row), helicity (middle row) and transversity (bottom row) PDFs for different values of $P_3$. Each sub-figure shows the momentum dependence of the corresponding distribution, where the blue line corresponds to $P_3=0.41\,{\rm GeV}$, the green line to $P_3=0.83\,{\rm GeV}$, and the red one to $P_3=1.24\,{\rm GeV}$.}
    \label{fig:vec_sca_momdep}
\end{figure}

In all cases, we find that the matrix elements for the lowest momentum $P_3=0.41$ GeV do not decayed to zero for large $z$, demonstrating, as expected, that the momentum is not large enough. By increasing the momentum to $P_3=1.24$~GeV, the matrix elements become consistent with zero within their uncertainties. While the imaginary parts show a residual momentum dependence, the convergence must be checked at the level of the reconstructed PDF. This is due to the fact that $P_3$ enters the matching kernel and  affects the convergence. 
Therefore, to address the momentum convergence as we increase $P_3$, we show in Fig.~\ref{fig:vec_sca_momdep} the momentum dependence of the isoscalar and isovector PDFs. We use the standard Fourier transform, with the values of $z_{\rm max}$ given in Table~\ref{table:cutoff_table}, as discussed in Sec.~\ref{sec:syst_eff_truncation}.
As can be seen in Fig.~\ref{fig:vec_sca_momdep}, the overall dependence on the two largest values of the  momentum is relatively small. Dependence on $P_3$ is observed in the unpolarized isoscalar PDF. In general, the PDFs for the smallest momentum, do not show convergence, and exhibit non-physical oscillations due to the presence of systematic effects in the reconstruction of the $x$-dependence. However, such oscillations are suppressed for the higher values of $P_3$.  
The isoscalar and isovector helicity distributions have a similar magnitude and exhibit milder dependence on  the boost as compared to the unpolarized. In particular, both isoscalar and isovector helicity  distributions are consistent for  $P_3=0.83\,\gev$ and $P_3=1.24\,\gev$. Finally, the isoscalar and isovector transversity distributions also show nice convergence with $P_3$ for the two largest values. These distributions will be used for the flavor decomposition  presented in Sec.~\ref{sec:flavor_dec} together with  comparison of our data with phenomenology.

\FloatBarrier

\subsection{Flavor decomposition and comparison with phenomenology}\label{sec:flavor_dec}

\subsubsection{Light quark distributions}\label{sec:light_quarks_distro}

Our results on the isoscalar and isovector distributions presented in Sec.~\ref{sec:vec_and_sca_distr} allow us to extract the up and down quark contributions for the unpolarized, helicity and transversity distributions. The disconnected contributions are taken into account in all cases. We stress that the comparison with phenomenology can only be qualitative for a number of reasons: i) We use an ensemble with larger than physical pion mass. We know from previous studies that there is a non-negligible pion mass dependence on the PDFs; ii) lattice systematics, such as cut-off effects, are not taken into account; iii) the renormalization ignores mixing present in the case of the unpolarized and helicity singlet PDFs; and iv) errors are still sizable and may hide systematics, such as convergence with the boost.  However, it is still interesting to compare with phenomenology keeping these caveats in mind. 
The results for the unpolarized PDF at the largest  momentum are compared with data by $\rm NNPDF3.1$~\cite{Ball:2017nwa}, while the helicity distribution is compared with JAM17~\cite{Ethier:2017zbq} and $\rm NNPDF_{\rm POL1.1}$~\cite{Nocera:2014gqa}. Finally, the quark transversity distribution obtained in this study is compared against the SIDIS data~\cite{Lin:2017stx} and SIDIS data constrained by the value of tensor charge $g_T$ computed in lattice QCD~\cite{Lin:2017stx}. For the anti-quark region for the $\rm NNPDF3.1$ data, we include the crossing relations of Eq.~\eqref{eq:crossing_rels}, such that we show the antiquark distributions in the negative-$x$ region.
\begin{figure}[ht!]
    \centering
    \includegraphics[width=0.9\linewidth]{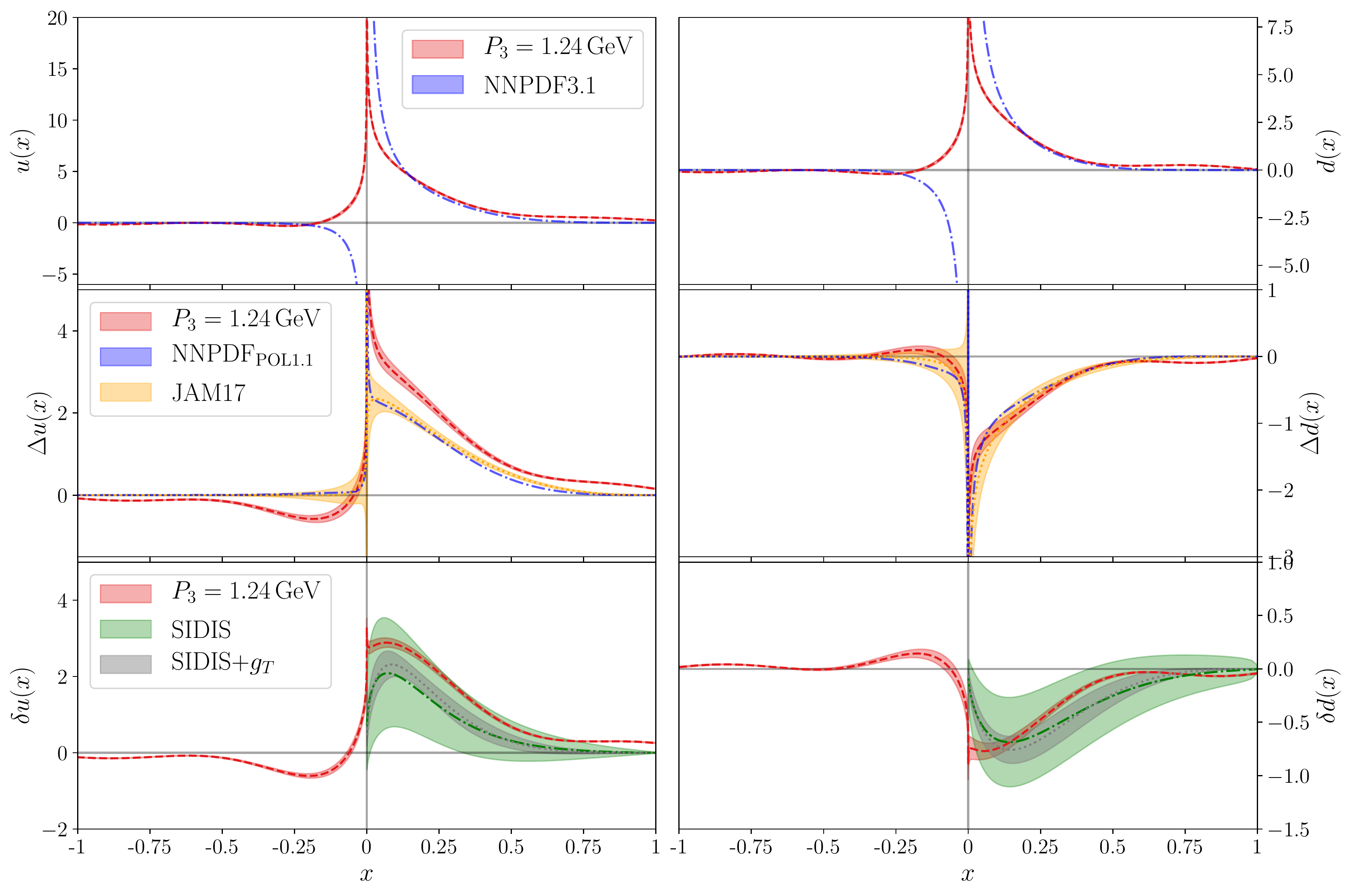}
    \caption{ Up (left) and down (right) quark unpolarized (upper panels), helicity (middle panels) and transversity (bottom panels) distributions at $P_3=1.24\,\gev$ (red band). We also show the $\rm NNPDF$ results~\cite{Ball:2017nwa,Nocera:2014gqa,Buckley:2014ana} (blue band) and JAM17~\cite{Ethier:2017zbq} (orange band) phenomenological results. For the transversity PDF we compare against the SIDIS data~\cite{Lin:2017stx} (green band) and SIDIS data constrained by the value of tensor charge $g_T$ computed in lattice QCD~\cite{Lin:2017stx} (gray band).}
    \label{fig:ud_distr}
\end{figure}

The light-quark contributions to the unpolarized PDF show good agreement with phenomenology in the region $x\gtrsim0.2$. Also, the region $x\lesssim-0.2$ both estimates are compatible with zero. Note that lattice results for the small-$x$ region ($|x|\lesssim 0.15$) suffer from uncontrolled uncertainties due to the reconstruction of the PDFs and the values of the lattice spacing used. The case of the helicity distributions is very interesting, as it has non-negligible contribution from the disconnected diagram.
Our results for the up quark helicity show similar features as the NNPDF data, but are have higher values. The down quark distribution gives compatible results both with $\rm NNPDF_{\rm POL1.1}$ and JAM17 data for all $x$ in the physical region $[-1,1]$. 
The transversity distribution is the least known collinear PDF and it is not well-constrained by SIDIS data. As a result, global fits for the light quark $\delta q(x)$ carry large relative error of $\approx50-100\%$ ~\cite{Lin:2017stx}. A more precise phenomenological estimate of the transversity PDFs can be obtained by constraining the distributions with the value of the tensor charge $g_T$ computed within lattice QCD~\cite{Lin:2017stx}.
A comparison with the latter, reveals a similar agreement as for the helicity PDFs. We would like to stress that the overall qualitative agreement is very promising, as this computation is done using simulations with heavier than physical pions.

\subsubsection{Strange quark distributions}\label{sec:strange_distr}

\begin{figure}[h!]
    \centering
    \includegraphics[width=0.7\linewidth]{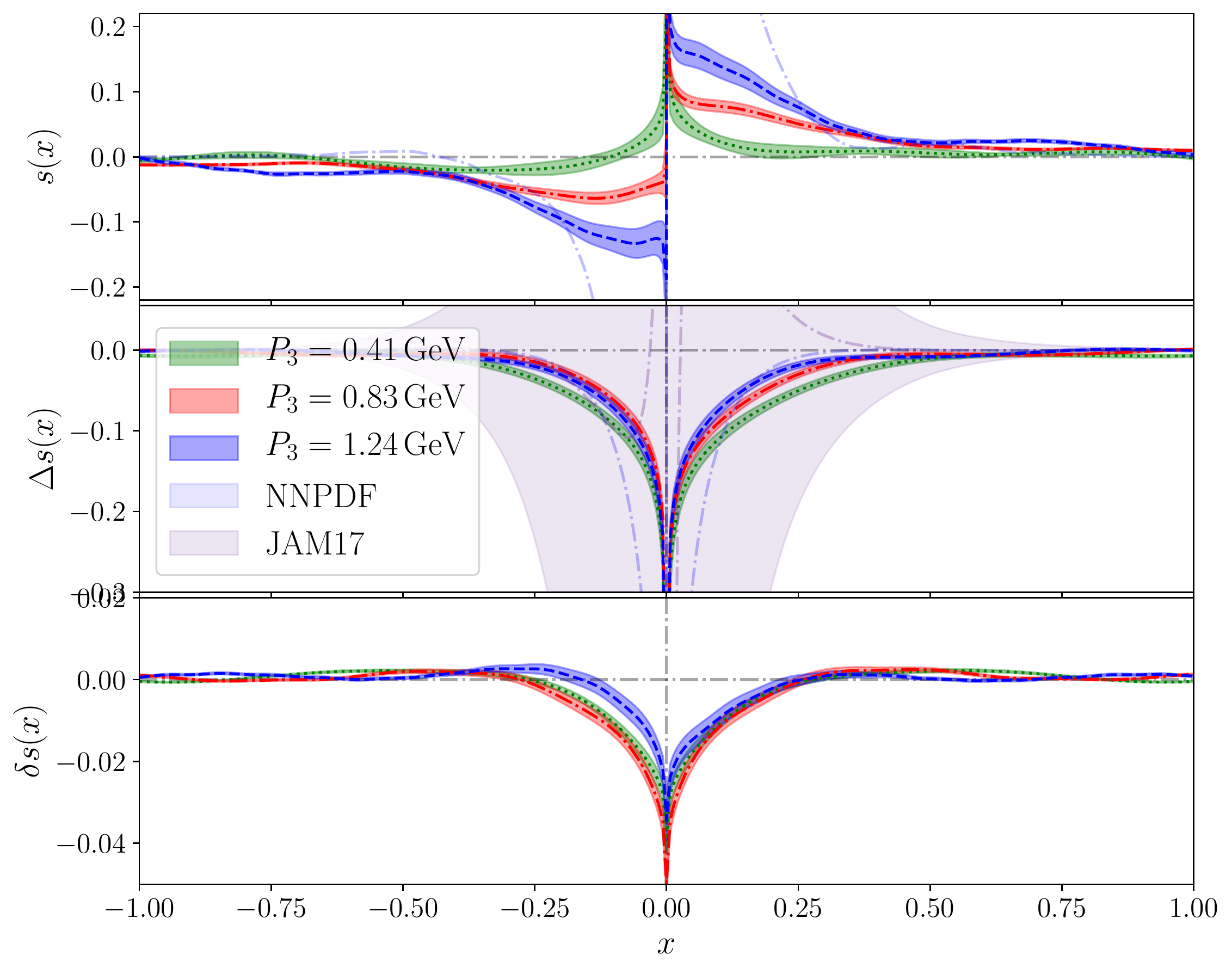}
    \caption{Results on the strange unpolarized (top panel), helicity (center panel) and transversity (bottom panel) distributions for three values of $P_3$. We compare with the  $\rm NNPDF_{\rm POL1.1}$~\cite{Buckley:2014ana,Nocera:2014gqa} (light blue) and JAM17~\cite{Ethier:2017zbq} (light purple) phenomenological data. Lattice data for $P_3=0.41,\,0.83,\,1.24$ GeV are shown with green, red and dark blue bands, respectively. }
    \label{fig:s_distr}
\end{figure}

The strange distributions presented here are computed using the renormalized matrix elements shown in Fig.~\ref{fig:mom_dep_str_me}. The values of $z_{\rm max}$ employed in the Fourier transform defining the quasi-PDF are reported in  Table~\ref{table:cutoff_table}. The criterion adopted to select $z_{\rm max}$  is to analyze the dependence of the PDF as $z_{\rm max}$ is increased, as discussed in the previous section. In Fig.~\ref{fig:s_distr} we show the unpolarized, helicity and transversity PDFs. The antiquark distribution reported here takes into account the crossing relations in Eq.~\eqref{eq:crossing_rels}, showing the anti-quark distributions in the negative $x$ region. Although the unpolarized PDFs extracted from the matrix element using the two largest momenta tend towards the phenomenological result, there is still some residual dependence, which points to the need to increase the momentum boost to check the independence on $P_3$. Due to the simultaneous suppression of the real part of the matrix elements and the enhancement of the imaginary part, $\bar{s}(x)$ becomes symmetrical with respect to $x=0$ as the momentum boost increases. This symmetry feature is exploited in the global fits.
The results for the helicity distribution are approximately symmetric in the quark and antiquark regions, and are compatible with the results from the $\rm NNPDF_{\rm POL1.1}$~\cite{Nocera:2014gqa} and with JAM17 global fits analysis both of which have larger uncertainties. Our results, thus, provide valuable input for phenomenological studies.
In fact, this is more evident for the strange transversity distribution where experimental results are lacking. We obtained  results on the  transversity PDF with small uncertainties that show no residual momentum dependence for the two largest momentum values. 

\begin{figure}[h!]
    \centering
    \includegraphics[width=0.6\linewidth]{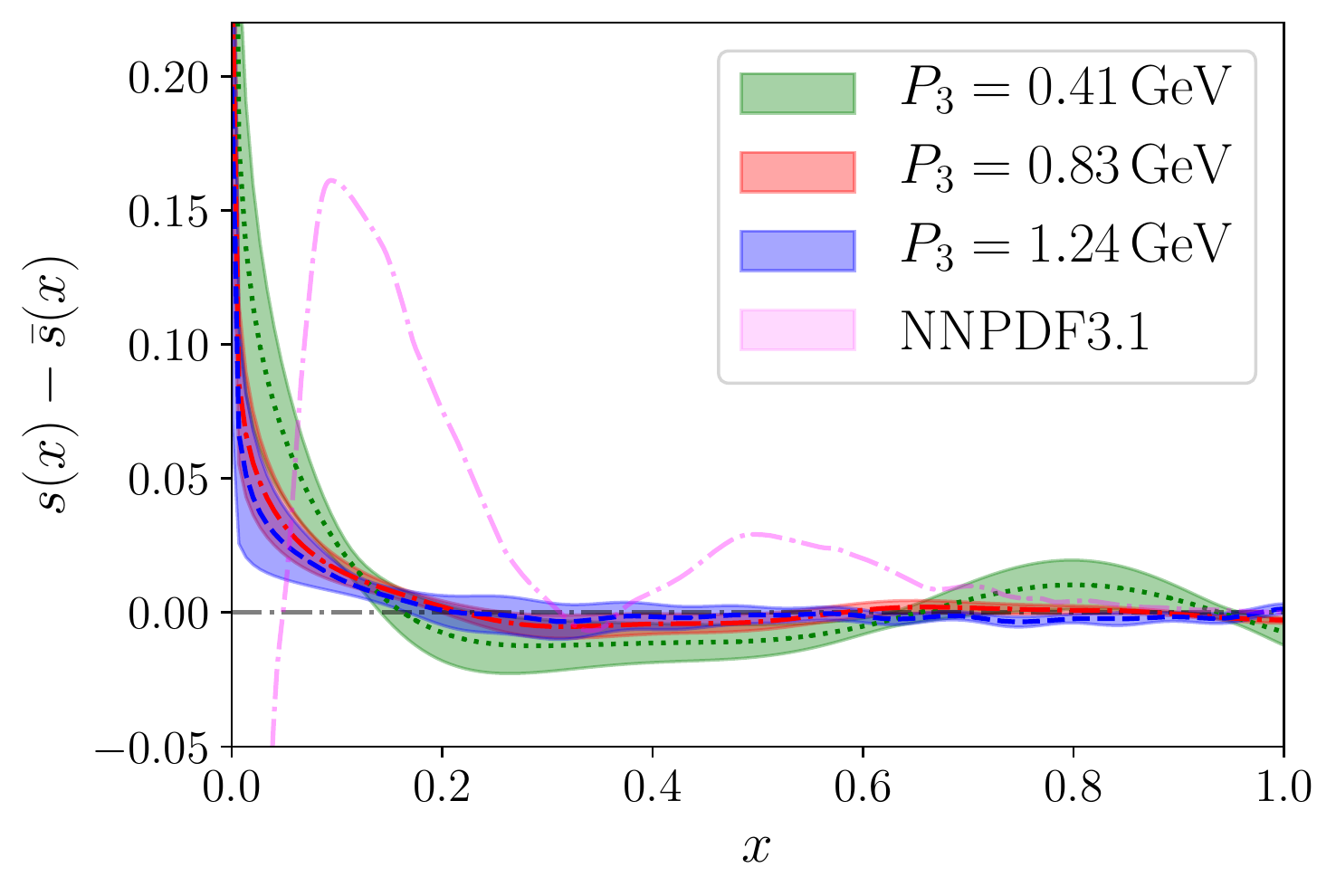}
    \caption{The strange-quark asymmetry for the unpolarized PDF for three values of $P_3$. We compare with $\rm NNPDF$~\cite{Buckley:2014ana} (pink) phenomenological data. Lattice data for $P_3=0.41,\,0.83,\,1.24$ GeV are shown with green, red and dark blue bands, respectively. }
    \label{fig:s_unpol_asymmetry}
\end{figure}

Besides the individual $s(x)$ and $\bar{s}(x)$ distributions, there is also an interest on the strange-quark asymmetry. This is partly due to the fact that there is no symmetry to suggest that the two distributions have to be the same. The strange and anti-strange asymmetry has been discussed within chiral effective theory~\cite{Wang:2016ndh,Wang:2016eoq}, perturbative evolution of QCD~\cite{Catani:2004nc}, and a physical model for parton momenta~\cite{Alwall:2004rd}. Here, we study the asymmetry using our data for $P_3=0.41,\,0.83,\,1.24$ GeV, and the results are shown in Fig.~\ref{fig:s_unpol_asymmetry}. In contrast to the individual $s(x)$ and $\bar{s}(x)$ distributions, here we find that there is no momentum dependence in the strange-quark asymmetry. We also note that the difference between $s(x)$ and $\bar{s}(x)$ is a non-singlet combination and, thus, does not mix with the gluon PDFs. Focusing on the most accurate results at $P_3$, we find that the asymmetry vanishes at $x\gtrsim 0.2$ and is small but non negligible in the small-$x$ region. This conclusion is, at present stage, qualitative, and an investigation of systematic effects is needed before drawing quantitative conclusions.

\FloatBarrier
\subsection{Moments of nucleon PDFs}
In this section, we calculate the  moments $\braket{x^n}$ of the three PDFs considering $n=0,...,3$. The $n-$th moment of the unpolarized, helicity and transversity distributions are defined as
\begin{equation}
\begin{split}
    \braket{x^n}_q &= \int_0^1 x^n\, \left[q(x)+(-1)^{n+1}\bar{q}(x)\right]\,dx = \int_{-1}^{1} x^n\,q(x)\,dx, \\
    \braket{x^n}_{\Delta q} &= \int_0^1 x^n\, \left[\Delta q(x)+(-1)^n\Delta \bar{q}(x)\right]\,dx =  \int_{-1}^{1} x^n\,\Delta q(x)\,dx, \\
    \braket{x^n}_{\delta q} &= \int_0^1 x^n\, \left[\delta q(x)+(-1)^{n+1}\delta \bar{q}(x)\right]\,dx = \int_{-1}^{1} x^n\,\delta q (x)\,dx,
\end{split}
\end{equation}
where we employed the crossing relations of Eq.~\eqref{eq:crossing_rels}, to write the moments as a function of the quark distributions only. In Table~\ref{table:moments_table} we report the results for the isovector, isoscalar and flavor diagonal moments. The zero-th moments are compatible with the nucleon charges reported in Tab~\ref{tab:res_charges}. This is a non-trivial check, as the calculation of the charges follows a totally different procedure and undergoes a Fourier transform and matching. 

\begin{table}[h!]
\begin{tabular}{c c | c c c c c } 
 \hline
 \hline
 \noalign{\vskip 0.1cm}    
    PDF &  & $u-d$ & $u+d$ & $u$ & $d$ & $s$\\ 
    \hline\hline
     Unpolarized & $\braket{x}_q$ & 0.28(1) & 0.75(2) & 0.51(2) & 0.234(9) & 0.030(2) \\
    & $\braket{x^2}_q$ & 0.118(5) & 0.23(1) & 0.176(8) & 0.058(4) & -0.00054(46) \\
    \hline
    Helicity & $\braket{1}_{\Delta q}$ & 1.26(6) & 0.50(6) & 0.88(5) & -0.38(3) & -0.033(3) \\
     & $\braket{x}_{\Delta q}$ & 0.49(2) & 0.32(2) & 0.40(2) & -0.087(9) & -0.00029(26) \\
    & $\braket{x^2}_{\Delta q}$ & 0.127(9) & 0.067(7) & 0.097(7) & -0.030(4) & -0.0019(4) \\
    \hline
    Transversity & $\braket{1}_{\delta q}$ & 1.06(4) & 0.67(4) & 0.86(3) & -0.20(2) & -0.0015(6) \\
     & $\braket{x}_{\delta q}$ & 0.49(2) & 0.33(2) & 0.41(2) & -0.075(7) & -0.00038(37) \\
    & $\braket{x^2}_{\delta q}$ & 0.118(6) & 0.086(5) & 0.102(5) & -0.016(2) & 0.00038(9) \\
  \noalign{\vskip 0.1cm}  
 \hline
 \hline
\end{tabular}
\caption{ Moments of the unpolarized, helicity and transversity PDFs. We refer to the zero-th moment  $\braket{x^0}$ as $\braket{1}$.}
\label{table:moments_table}
\end{table}

\FloatBarrier
\section{Conclusions}
\label{sec:concl}

In this work we present a study of the $x$-dependence of proton collinear quark PDFs from lattice QCD considering both connected and disconnected diagrams. These contributions are necessary to determine the individual-flavor contributions to PDFs. We present results for the up, down and strange quark for the unpolarized, helicity and transversity PDFs. This work extends our first calculation for the flavor decomposition of the helicity PDFs~\cite{Alexandrou:2020uyt}; here we increase the statistics by about a factor of two, and include results on the unpolarized and transversity PDFs. 

The main goal of this work is to explore the feasibility of the calculation of disconnected quark loops with non-local operators. To this end, we provide the necessary details on technical and theoretical aspects, as well as the examination of some sources of systematic uncertainties. 
The calculation is carried out using one ensemble of $N_f=2+1+1$ twisted mass fermions simulated with quark mass value that produces a pion mass of 260 MeV. Using a single ensemble, we can address excited-states contamination, reconstruction of the $x$ dependence, and the convergence with increasing the momentum boost in the final PDFs. 

The matrix elements contain non-local operators with the length of the Wilson line extending up to half the spatial extend of the lattice. The proton states are boosted with momentum using three values, namely $P_3=0.41,\,0.83,\,1.24$ GeV. Several values of the source-sink time separation are considered.  As we  increase the boost  we also increase the source-sink time separation in order to investigate the effect of excited states; we employ up to $t_s=1.13$ fm for the highest momentum (see Figs.~\ref{fig:summary_z3_exc_st_unp} - \ref{fig:comparison_m3_diff_meths}). Both the isovector and isoscalar flavor combinations are calculated, with the latter receiving contributions from the connected and disconnected diagrams. All matrix elements are renormalized multiplicatively using the RI$'$ scheme and evolved to the modified-$\MSbar$ scheme at a scale of 2 GeV (unpolarized and helicity) or $\sqrt{2}$ GeV (transversity). The renormalization is followed by the transform to the momentum space, $x$, which produces the quasi-PDFs. We apply the standard Fourier transform that is confirmed using the Bayes-Gauss-Fourier transform, finding compatible results (see Fig.~\ref{fig:zmax_dep_vecsca}). The quasi-PDFs are matched to the light-cone PDFs using one-loop perturbation theory. The matching kernel contains information on the renormalization scheme and scale for the quasi-PDFs (modified-$\MSbar$ scheme at 2 or $\sqrt{2}$ GeV), which are then matched to the light-cone PDFs in the $\MSbar$ scheme at the same scale. In this proof-of-principle study we neglect the mixing with the gluon PDFs for the unpolarized and helicity case. The extraction of the latter has its own challenges and will be included in our future studies. To test the influence of systematic uncertainties, we calculate the charges by integrating the PDFs (Table~\ref{tab:res_charges}), and compare with the values obtained directly from the matrix elements (Table~\ref{table:moments_table}). We find good consistence among the results.

We find that the light-quark disconnected contributions have the most impact for the helicity PDF, while the transversity disconnected contribution is very small. Regardless, a clear non-zero signal is found in all cases. The strange-quark PDFs are nonzero up to $x\sim 0.5$, with the unpolarized and helicity  having a similar magnitude, and the transversity being an order of magnitude smaller, as can be seen in Fig.~\ref{fig:s_distr}. These distributions are very challenging to extract from experimental data due to the lack of sensitivity to the strange-quark. In a qualitative comparison of our results with phenomelogically extracted PDFs we find: i) our results on the unpolarized have a statistical precision which is similar to the NNPDF data; ii) the helicity strange-quark PDF is significantly more accurate than the JAM and NNPDF results and iii)  our results for the strange-quark transversity PDF serve as a prediction.

There are a number of improvements that one can do to quantify and eliminate systematic uncertainties. These include, but not limited to, pion mass dependence, mixing under matching with the gluon PDFs for the unpolarized and helicity case, and address the inverse problem using, e.g., Bayesian reconstruction methods. One can also address finite-volume and discretization effects, which require extracting the PDFs using multiple ensembles.  However, this work clearly demonstrates the great potential in the extraction of the x-dependence of individual quark PDFs from lattice QCD. 

\FloatBarrier
\begin{acknowledgments}
We would like to thank all members of ETMC for their constant and pleasant collaboration. We also thank Fernanda Steffens and Jeremy Green for useful discussions.
M.C. acknowledge financial support by the U.S. Department of Energy Early Career Award under Grant No.\ DE-SC0020405.
  K.H. is supported financially by the Cyprus Research and  Innovation Foundation under contract number POST-DOC/0718/0100. 
  This project has received funding from the Marie Skłodowska-Curie
  European Joint Doctorate program STIMULATE of the European
  Commission under grant agreement No 765048; F.M. is funded under
  this program.
  This research includes calculations carried out on  HPC resources of the Cyprus Institute  (the  Cyclone and Cyclamen machines) and  of Temple University, supported in part by the National Science Foundation through major research instrumentation grant number 1625061 and by the US Army Research Laboratory under contract number W911NF-16-2-0189. Computations for this work were carried out in part on facilities of the USQCD Collaboration, which are funded by the Office of Science of the U.S. Department of Energy. This research used resources of the Oak Ridge Leadership Computing Facility, which is a DOE Office of Science User Facility supported under Contract DE-AC05-00OR22725. The gauge configurations have been generated by the Extended Twisted Mass Collaboration on the KNL (A2) Partition of Marconi at CINECA, through the Prace project Pra13\_3304 "SIMPHYS". 
\end{acknowledgments}

\appendix

\bibliography{pdf_bib.bib}

\end{document}